\documentclass{article}
\usepackage[top=0.75in, bottom=0.75in, left=0.75in, right=0.75in]{geometry} 
\usepackage{amssymb}
\usepackage{amsmath}
\usepackage{graphicx}
\usepackage[numbers,sort&compress]{natbib}
\usepackage{mathtools}
\usepackage{booktabs}
\usepackage{siunitx}
\usepackage{subcaption}
\usepackage{multirow}
\usepackage{setspace}
\usepackage{hyperref}
\usepackage[percent]{overpic} 

\newcommand\doubleplus{+\kern-1.3ex+}  

\newcommand{\textb}[1]{#1} 

\renewcommand{\eqref}[1]{Equation~(\ref{eq:#1})}

\newcommand{\tabref}[1]{Table~\ref{tab:#1}} 
\newcommand{\figref}[1]{Figure~\ref{fig:#1}} 

\DeclareMathOperator*{\argmin}{arg\,min}

\renewcommand{\Tilde}{\widetilde}
\renewcommand{\Hat}{\widehat}

\renewcommand{\vec}[1]{\ensuremath{\boldsymbol{#1}}}
\newcommand{\Tvec}[1]{\ensuremath{\Tilde{\boldsymbol{#1}}}}
\newcommand{\Hvec}[1]{\ensuremath{\Hat{\boldsymbol{#1}}}}

\newcommand{\Real}{{\mathbb{R}}}
\newcommand{\Complex}{{\mathbb{C}}}
\newcommand{\dvec}[1]{\ensuremath{\Dot{\boldsymbol{#1}}}}
\newcommand{\ddvec}[1]{\ensuremath{\Ddot{\boldsymbol{#1}}}}

\newcommand{\footremember}[2]{%
    \footnote{#2}
    \newcounter{#1}
    \setcounter{#1}{\value{footnote}}%
}
\newcommand{\footrecall}[1]{%
    \footnotemark[\value{#1}]%
}
\let\svthefootnote\thefootnote
\newcommand\freefootnote[1]{%
  \let\thefootnote\relax%
  \footnotetext{#1}%
  \let\thefootnote\svthefootnote%
}
\title{A multi-dynamic low-rank deep image prior (ML-DIP) for \textb{3D real-time} cardiovascular MRI}
\author{%
    Chong Chen\footremember{OSUbme}{Department of Biomedical Engineering, The Ohio State University, Columbus, OH, USA}%
    \and Marc Vornehm\footremember{FAU}{Department Artificial Intelligence in Biomedical Engineering, FAU Erlangen-N\"urnberg, Erlangen, Germany}\footremember{SHS}{Research \& 
    Clinical Translation, Magnetic Resonance, Siemens Healthineers AG, Erlangen, Germany}%
    \and \textb{Zhenyu Bu}\footrecall{OSUbme}%
    \and Preethi Chandrasekaran\footremember{DLHRI}{Davis Heart and Lung Research Institute, The Ohio State University, Columbus, OH, USA}%
    \and Muhammad A. Sultan\footrecall{OSUbme}%
    \and Syed M. Arshad\footremember{OSUece}{Department of Electrical and Computer Engineering, The Ohio State University, Columbus, OH, USA}%
    \and Yingmin Liu\footrecall{DLHRI}%
    \and Yuchi Han\footremember{OSUMC}{Division of Cardiovascular Medicine, The Ohio State University, Columbus, OH, USA}%
    \and Rizwan Ahmad\footrecall{OSUbme}~\footrecall{OSUece}%
}
\date{}

\begin{document}

\maketitle
\freefootnote{}
\freefootnote{Corresponding author: Rizwan Ahmad (ahmad.46@osu.edu)}

\begin{center}
\vspace{5mm}
\end{center}

\begin{abstract}
\noindent\textbf{Purpose:} To develop a reconstruction framework for 3D real-time cine cardiovascular magnetic resonance (CMR) from highly undersampled data without requiring fully sampled training datasets.

\noindent\textbf{Methods:} We developed a multi-dynamic low-rank deep image prior (ML-DIP) framework that models spatial image content and deformation fields using separate neural networks. These sub-networks are jointly trained per scan to reconstruct the dynamic image series directly from undersampled k-space data. ML-DIP was evaluated on (i) a 3D cine digital phantom with simulated premature ventricular contractions (PVCs), (ii) ten healthy subjects (including two scanned during both rest and exercise), and (iii) \textb{12 patients with a history of PVCs.} Phantom results were assessed using peak signal-to-noise ratio (PSNR) and structural similarity index measure (SSIM). In vivo performance was evaluated by comparing left-ventricular function quantification (against 2D real-time cine) and image quality (against 2D real-time cine and binning-based 5D-Cine).

\noindent\textbf{Results:} In the phantom study, ML-DIP achieved PSNR $>29$ dB and SSIM $>0.90$ for scan times as short as two minutes, while recovering cardiac motion, respiratory motion, and PVC events. In healthy subjects, ML-DIP yielded functional measurements comparable to 2D cine and higher image quality than 5D-Cine, including during exercise with high heart rates and bulk motion. In PVC patients, ML-DIP preserved beat-to-beat variations and reconstructed irregular beats, whereas 5D-Cine showed motion artifacts and information loss due to binning.

\noindent\textbf{Conclusion:} ML-DIP enables high-quality 3D real-time CMR with acceleration factors exceeding $1,\!000$ by learning low-rank spatial and motion representations from undersampled data, without relying on external fully sampled training datasets.
\end{abstract}

\subsection*{Abbreviations}
2D, two-dimensional; 3D, three-dimensional; 4D, four-dimensional; 5D, five-dimensional; 5D-Cine, cardiorespiratory-binning-based volumetric cine; BMI, body mass index; BSA, body surface area; CMR, cardiovascular magnetic resonance imaging; CNN, convolutional neural network; DIP, deep image prior; ECG, electrocardiogram; EDV, end-diastolic volume; EF, ejection fraction; ESPIRiT, eigenvalue approach to autocalibrating parallel MRI; ESV, end-systolic volume; FFT, fast Fourier transform; LR-DIP, low-rank deep image prior; LV, left ventricle; M-DIP, multi-dynamic deep image prior; ML-DIP, multi-dynamic low rank deep image prior; MoCo-SToRM, motion-compensated smoothness regularization on manifolds; MRXCAT, magnetic resonance extended cardiac-torso; OPRA, ordered pseudo-radial sampling; PCA, principal component analysis; PSNR, peak signal-to-noise ratio; PVC, premature ventricular contraction; R, acceleration rate; SI, superior-inferior; SSIM, structural similarity index measure; SV, stroke volume; XCAT, extended cardiac-torso; 

\section{Introduction}
Cardiovascular magnetic resonance imaging (CMR) is a well-established diagnostic imaging modality. CMR data are often collected slice-by-slice with electrocardiogram (ECG) gating and during breath-holds. This approach fails in patients who cannot hold their breath or have arrhythmia. For such subjects, free-breathing real-time imaging is used as a fallback option. Two-dimensional (2D) real-time imaging has progressed at a rapid pace over the past two decades \cite{contijoch2024future}, with advanced acceleration techniques enabling spatial and temporal resolutions comparable to breath-held segmented acquisitions. However, it remains limited in visualizing and modeling 3D structures due to through-plane motion, slice misregistration, and a slice thickness of 5 to 8 mm. Although 3D imaging with volumetric coverage offers advantages over 2D imaging, existing 3D imaging paradigms have significant limitations of their own.
In particular, 3D methods that use prospective gating suffer from unpredictable and prolonged scan times \cite{goo2018comparison}, while self-gating approaches that mitigate this issue \cite{feng20185d} remain sensitive to binning errors, which can degrade image quality \cite{arshad2024motion}.
For patients with arrhythmias or irregular respiratory patterns, the image quality is often compromised due to intra-bin and inter-bin motion.
Even when successful, these 3D methods cannot image beat-to-beat variations, which may carry diagnostic and prognostic value \cite{dicarlo2023assessment, alhede2022premature}, but instead reconstruct one ``typical'' cardiac and/or respiratory cycle.

Extending real-time imaging to 3D would circumvent the limitations of 2D real-time imaging and 3D binning-based imaging but requires extremely high acceleration rates ($R>500$). One of the few efforts in this direction is the 2023 work by Sun et al., who proposed motion-resolved real-time 4D flow imaging using low-rank (LR) and subspace modeling and applied it to study beat-to-beat flow variations in ten healthy subjects and two patients \cite{sun2023motion}. However, this approach requires a long acquisition time of 12.18$\pm$1.39 minutes for imaging the aorta alone. Furthermore, depending on the selected rank and matrix size, the memory requirements of this method can be prohibitive. \textb{Earlier, in 2021, Huttinga et al. introduced MR-MOTUS, which estimates respiratory and cardiac motion fields directly from undersampled data via a low-dimensional parameterization of deformation, and demonstrated its feasibility for real-time MR-guided radiotherapy \cite{huttinga2021nonrigid}.} In 2022, Zou et al.\ presented motion-compensated smoothness regularization on manifolds (MoCo-SToRM) for lung imaging \cite{zou2022dynamic}. In this work, they approximated all frames in the time series as respiration-deformed versions of a single 3D template image. The deformation fields were modeled as output of a convolutional neural network (CNN) driven by low-dimensional latent vectors. Both the CNN weights and the 3D template image were jointly estimated from the undersampled k-space data. Recently, Kettelkamp et al.\ proposed DMoCo, which models 3D cardiac MRI volumes across motion phases as diffeomorphic deformations of a single template, and provided proof-of-concept results with radial sampling and 70\,ms temporal resolution \cite{kettelkamp2023motion,kettelkamp2025diffeomorphic}. However, this framework also uses a single template image and thus cannot handle contrast fluctuations, which are inevitable, e.g., due to inflow enhancement. Hamilton et al.\ proposed a low-rank deep image prior (LR-DIP) framework for 2D real-time imaging and then extended it to 3D real-time imaging \cite{hamilton20243d} using a stack-of-spirals acquisition \textb{with anisotropic resolution.} However, LR-DIP captures motion implicitly through low-rank modeling, \textb {whereas our recent 2D work} shows that explicit motion modeling improves reconstruction quality and generates sharper images \cite{vornehm2025multi}.


In this work, we propose and evaluate the multi-dynamic low-rank deep image prior (ML-DIP) framework, which models both motion and content variation using low-rank representations of the image and deformation field. Building on our prior 2D M-DIP framework \cite{vornehm2025multi,vornehm2025motionguided}, we add low-rank modeling of deformation fields, which enables 3D imaging without requiring the memory footprint of the convolutional decoder to scale with batch size. Unlike LR-DIP \cite{hamilton20243d} and the method by Sun et al. \cite{sun2023motion}, ML-DIP explicitly models motion using deformation fields, making it more constrained. In contrast to the work by Kettelkamp et al. \cite{kettelkamp2023motion,kettelkamp2025diffeomorphic}, which directly solves for the image template and deformation basis with explicit spatial regularization, ML-DIP models both the image and deformation bases as outputs of CNNs and thus benefits from their inductive bias as an implicit regularizer. More importantly, ML-DIP models image content variation via a trainable image basis, facilitating recovery of dynamic changes beyond motion alone. These features make ML-DIP well-suited for a wide range of 3D real-time applications where both motion and contrast evolve across frames.

\section{Methods}
\subsection{Real-time imaging in 3D}
In 3D real-time imaging, the goal is to recover a series of images, each of size $n_1\times n_2\times n_3$. Let $\vec{x}^{(1:T)} \coloneq \big\{\vec{x}^{(t)}\big\}_{t=1}^{T}$ represent the 3D image series, where $T$ is the total number of frames and $\vec{x}^{(t)}\in\Complex^{N\times 1}$ is the vectorized version of the $t^{\text{th}}$ frame with $N = n_1 \times n_2 \times n_3$ voxels. Throughout this paper, we will use the notation $(\cdot)^{(1:T)}$ to represent time series of arrays or operations with $T$ frames. \textb{For an acquisition with $M$ measured k-space samples per frame, let $\vec{y}^{(t)}\in\Complex^{M\times 1}$, $\vec{A}^{(t)}\in\Complex^{M\times N}$, and $\vec{\epsilon}^{(t)}\in\Complex^{M\times 1}$ represent the noisy multi-coil k-space data, forward operator, and additive white Gaussian noise with variance $\sigma^2$, respectively, for the $t^{\text{th}}$ frame.} One could attempt to solve this problem using regularized least squares, i.e.,

\begin{equation}
    \Hvec{x}^{(1:T)} = \argmin _{\vec{x}^{(1:T)}}\sum_{t=1}^T\left\|\vec{A}^{(t)}\vec{x}^{(t)} - \vec{y}^{(t)}\right\|_2^2 + \textb{\lambda}\mathcal{R}(\vec{x}^{(1:T)}),
    \label{eq:ls}
\end{equation}
where the term $\mathcal{R}(\vec{x}^{(1:T)})$ represents spatial and/or temporal regularization \textb{controlled by $\lambda\geq0$}. However, due to the extremely high acceleration rates and large memory demands of 3D real-time imaging, directly solving the $\vec{y}^{(t)} = \vec{A}^{(t)}\vec{x}^{(t)} + \vec{\epsilon}^{(t)}$ problem is generally infeasible. Binning-based recovery methods address this by distributing the collected k-space data into a discrete number of cardiorespiratory bins and solving the problem in \eqref{ls} to reconstruct a representative cardiac and/or respiratory cycle. While these methods have been successful in many research settings \cite{ma20205d, di2019automated, roy2022free}, they are not real-time and cannot capture beat-to-beat variations. Additionally, these methods can degrade significantly or fail completely in the presence of frequent arrhythmias, inconsistent respiratory motion, or bulk motion.

\subsubsection{Extending deep image prior to 3D imaging}
Deep image prior (DIP) provides an unsupervised learning framework for solving inverse problems without requiring training data \cite{ulyanov2018deep}. 
In DIP, a generative network is trained to map a random code vector to an output consistent with the measurements.
A key feature of DIP is that the network structure acts as an implicit prior, eliminating the need for explicit regularization. Another key feature of DIP is that it is instance-specific, i.e., the network training is performed from scratch for each set of measurements.

A natural extension of DIP for dynamic imaging involves combining it with manifold learning. This is achieved by modeling the nonlinear mapping using a neural network:
\begin{equation}
    \vec{x}^{(t)} = \mathcal{G}_{\vec{\xi}} (\vec{z}^{(t)}),
    \label{eq:nonlinear}
\end{equation}
\textb{where $\mathcal{G}_{\vec{\xi}} \colon \Real^{K\times 1} \rightarrow \Complex^{N\times 1}$ is a network parameterized by $\vec{\xi}$ and $\vec{z}^{(t)} \in \Real^{K\times 1}$ is a low-dimensional latent code vector of user-defined dimensionality $K$.} 

For given multi-coil k-space data ~$\vec{y}^{(t)}$ and forward operator $\vec{A}^{(t)}$, injecting the nonlinear mapping in \eqref{nonlinear} to the data consistency term in \eqref{ls} leads to this optimization problem:
\begin{equation}
     \Hvec{\xi},\Hvec{z}^{(1:T)} = \argmin_{\vec{\xi}, \vec{z}^{(1:T)}}\sum_{t=1}^T \left\| \vec{A}^{(t)} \mathcal{G}_{\vec{\xi}}(\vec{z}^{(t)}) - \vec{y}^{(t)} \right\|_2^2.
\end{equation}

After training, an arbitrary $t^{\text{th}}$ frame can be recovered by $\Hvec{x}^{(t)} = \mathcal{G}_{\Hvec{\xi}}(\Hvec{z}^{(t)})$.
Recently, several studies have used DIP to learn manifolds for 2D dynamic MRI applications \cite{yoo2021timedependent,zou2021dynamic}. While these approaches capture redundancy across frames through a shared network, they do not fully
capture the temporal structure in a dynamic image series $\vec{x}^{(1:T)}$. Since these approaches are not adequately constrained, our initial efforts to directly extend them to 3D real-time imaging, where the acceleration rates can be two orders of magnitude higher, were unsuccessful.

Recently proposed \textb{MoCo-SToRM \cite{zou2022dynamic}} offers a more constrained approach to extending DIP to 3D real-time imaging. Instead of directly generating the more complex $\vec{x}^{(1:T)}$, it uses the network $\mathcal{G}_{\vec{\xi}}$ to generate frame-specific 3-directional deformation fields $\vec{\phi}^{(t)} \in \Real^{N \times 3}$ and then solves the following optimization problem:

\begin{equation}
     \Hvec{\xi},\Hvec{z}^{(1:T)}, \Hvec{x}_{\text{st}} = \argmin_{\vec{\xi}, \vec{z}^{(1:T)}, \vec{x}_{\text{st}}}\sum_{t=1}^T \left\| \vec{A}^{(t)} \big(\vec{x}_{\text{st}} \circ \vec{\phi}^{(t)}\big) - \vec{y}^{(t)} \right\|_2^2, ~~~ \vec{\phi}^{(t)} = \mathcal{G}_{\vec{\xi}} (\vec{z}^{(t)}),
     \label{eq:mocostorm}
\end{equation}
where $\vec{x}_{\text{st}} \in \Complex^{N \times 1}$ denotes a single static template image and ``$\circ$'' denotes the spatial warping operation \cite{jaderberg2015spatial}. This framework, however, can only model motion but not other dynamics, such as contrast fluctuations. Also, this method infers $\vec{x}_{\text{st}}$ directly and does not model it as an output of a CNN. Therefore, it does not leverage the inductive bias of CNNs but rather relies on explicit regularization of $\vec{x}_{\text{st}}$, which is omitted from \eqref{mocostorm} for simplicity. The proposed method, described next, circumvents these limitations. 






\subsubsection{ML-DIP framework}

\begin{figure}[ht]
    \centering
    \includegraphics[width=\textwidth]{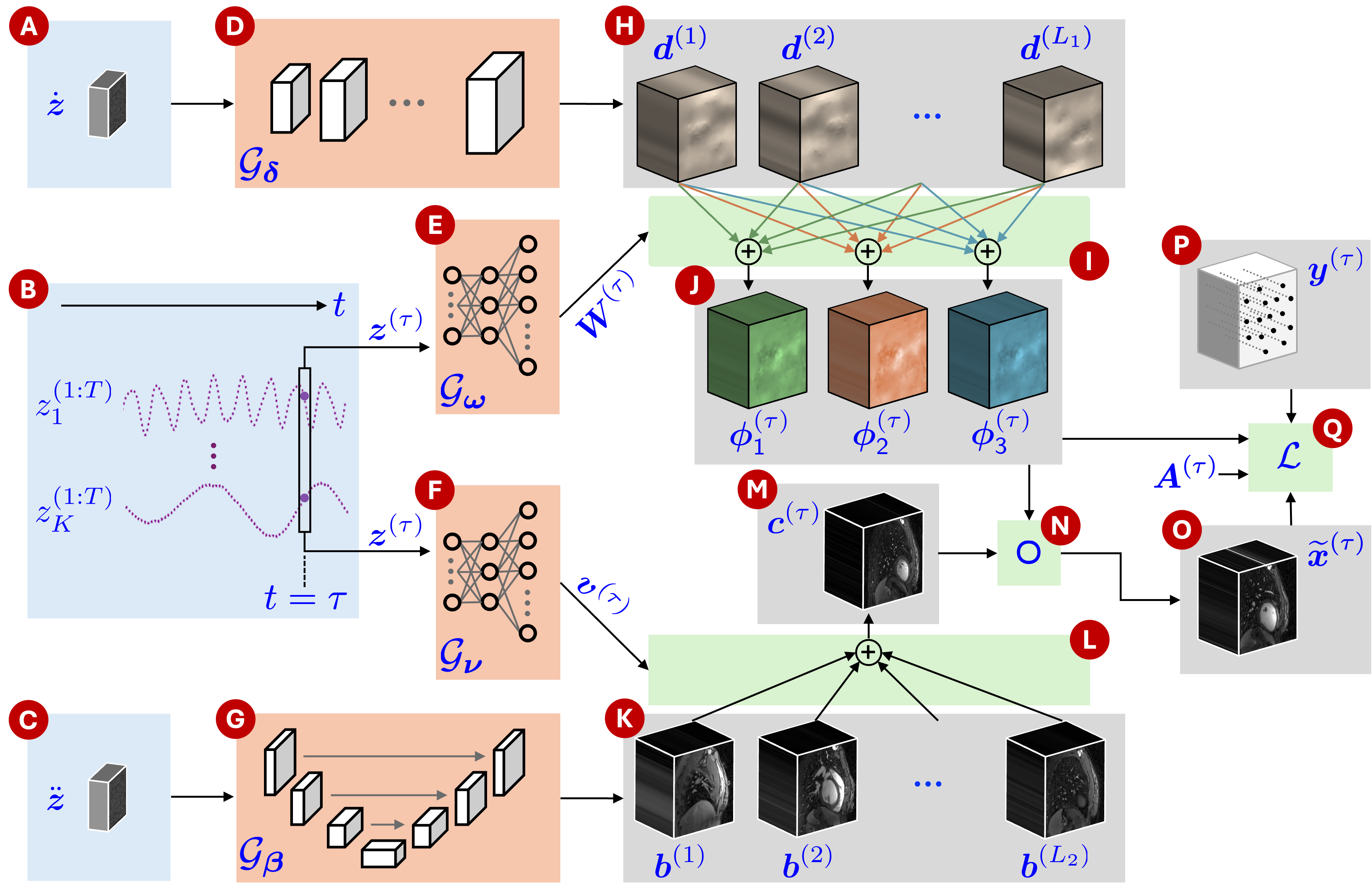}
    \caption{Overview of ML-DIP, showing the flow of information for the $\tau^{\text{th}}$ frame. A: Trainable static code vectors $\dvec{z}$ that serve as input to the generator $\mathcal{G}_{\vec{\delta}}$. B: A series of trainable dynamic code vectors $\vec{z}^{(1:T)}$, with a frame-specific code vector $\vec{z}^{(\tau)}$ serving as input to the generators $\mathcal{G}_{\vec{\omega}}$ and $\mathcal{G}_{\vec{\nu}}$. C: Trainable static code vectors $\ddvec{z}$ that serve as input to the generator $\mathcal{G}_{\vec{\beta}}$. D: Decoder-based CNN $\mathcal{G}_{\vec{\delta}}$ to generate deformation field basis. E: Fully connected network $\mathcal{G}_{\vec{\omega}}$ to generate frame-specific weights $\vec{W}^{(\tau)}$ to combine the elements of the deformation field basis. F: Fully connected network $\mathcal{G}_{\vec{\nu}}$ to generate frame-specific weights $\vec{v}^{(\tau)}$ for combining the elements of the \textb{image} basis. G: U-Net CNN $\mathcal{G}_{\vec{\beta}}$ to generate image basis. H: Static deformation field basis generated by $\mathcal{G}_{\vec{\delta}}$. I: Linearly combining the deformation basis elements using $\vec{W}^{(\tau)}$ to generate a frame-specific deformation field. J: The three components of the deformation field $\vec{\phi}^{(\tau)}$. K: Static image basis generated by $\mathcal{G}_{\vec{\beta}}$. L: Linearly combining the image basis elements using $\vec{v}^{(\tau)}$ to generate a frame-specific composite image. M: The complex-valued composite image $\vec{c}^{(\tau)}$. N: Warping operation ``$\circ$'' where the composite image $\vec{c}^{(\tau)}$ is deformed by $\vec{\phi}^{(\tau)}$. O: The predicted 3D frame $\Tvec{x}^{(\tau)}$. P: The measured multi-coil k-space data $\vec{y}^{(\tau)}$. Q: The loss function that penalizes the discrepancy between $\vec{A}^{(\tau)}\Tvec{x}^{(\tau)}$ and $\vec{y}^{(\tau)}$ and promotes spatial smoothness in the deformation fields. 
    }
    \label{fig:overview}
\end{figure}

ML-DIP integrates manifold learning, deformation-based motion modeling, and scalable low-rank representation into a unified framework. A high-level description of ML-DIP is provided in \figref{overview}. ML-DIP generates a deformation field basis and an image basis using two separate CNNs. The corresponding elements from these bases are then combined to yield a frame-specific deformation field and a frame-specific composite image. The composite image is subsequently warped using the deformation field to produce an output frame. This strategy confines the dynamic component of learning to a small set of weights used to combine the basis elements. Moreover, the low-rank representation of both the deformation field and the image enables modeling of motion and image content variations (e.g., contrast), while also reducing the size of the CNN generators. Finally, since the combination weights are generated by fully connected networks with a low-dimensional input, they support learning motion and content variations through manifold learning.

In ML-DIP, we jointly train four sub-networks and three sets of code vectors to generate an output frame that is consistent with the undersampled k-space data for that frame. As shown in \figref{overview}, a CNN-based generator $\mathcal{G}_{\vec{\delta}}$ takes a static code vector $\dvec{z}$ as input and generates deformation field basis $\vec{d}^{(1:L_1)}$ as output, where $L_1$ is the number of elements in the basis and $\vec{d}^{(i)}\in\Real^{N\times 1}$ is the $i^{\text{th}}$ element of the deformation basis. This generator uses a decoder architecture and is parameterized by $\vec{\delta}$. We refer to this network as ConvDecoder. Another CNN-based generator $\mathcal{G}_{\vec{\beta}}$ takes a different static code vector $\ddvec{z}$ as input and generates image basis $\vec{b}^{(1:L_2)}$ as output, where $L_2$ is the number of elements in the basis and $\vec{b}^{(i)}\in\Complex^{N\times 1}$ is the $i^{\text{th}}$ element of the image basis. This generator uses a U-Net architecture and is parameterized by $\vec{\beta}$. The outputs of $\mathcal{G}_{\vec{\delta}}$ and $\mathcal{G}_{\vec{\beta}}$ are static, i.e., a single deformation field basis and a single image basis are generated for the entire real-time image series. A small fully connected network $\mathcal{G}_{\vec{\omega}}$, parameterized by $\vec{\omega}$, takes frame-specific code vector $\vec{z}^{(t)}\in\Real^{K\times 1}$ of dimensionality $K$ as input and generates a frame-specific weight matrix $\vec{W}^{(t)} \in \Real^{L_1 \times 3}$. The role of $\vec{W}^{(t)}$ is to linearly combine elements of $\vec{d}^{(1:L_1)}$ into the frame-specific deformation field $\vec{\phi}^{(t)}=[\vec{\phi}_1^{(t)},\vec{\phi}_2^{(t)},\vec{\phi}_3^{(t)}]$, which is comprised of three components corresponding to the three spatial axes. The operation of $\vec{W}^{(t)}$ on $\vec{d}^{(1:L_1)}$ can be described in terms of matrix-matrix multiplication $\vec{D}\vec{W}^{(t)}$, where the matrix $\vec{D}\in\Real^{N\times L_1}$ contains elements of $\vec{d}^{(1:L_1)}$ as its columns. Another small fully connected network $\mathcal{G}_{\vec{\nu}}$, parameterized by $\vec{\nu}$, takes the same frame-specific $\vec{z}^{(t)}\in\Real^{K\times 1}$ as input and generates a frame-specific weight vector $\vec{v}^{(t)} \in \Complex^{L_2 \times 1}$. The role of $\vec{v}^{(t)}$ is to linearly combine elements of $\vec{b}^{(1:L_2)}$ into the frame-specific composite image $\vec{c}^{(t)}$. The operation of $\vec{v}^{(t)}$ on $\vec{b}^{(1:L_1)}$ can be described in terms of matrix-vector multiplication $\vec{B}\vec{v}^{(t)}$, where the matrix $\vec{B}\in\Complex^{N\times L_2}$ contains elements of $\vec{b}^{(1:L_2)}$ as its columns. The resulting frame-specific deformation field $\vec{\phi}^{(t)}$ is then used to spatially warp the frame-specific composite image $\vec{c}^{(t)}$ to generate $\Tvec{x}^{(t)}$, which is the prediction of the $t^{\text{th}}$ frame.

The joint training of the four sub-networks and three code vectors is realized by solving the following optimization problem.
\begin{equation}
    \Hvec{\delta}, \Hvec{\beta}, \Hvec{\omega}, \Hvec{\nu}, \Hat{\dvec{z}}, \Hat{\ddvec{z}}, \Hvec{z}^{(1:T)} = \argmin_{\vec{\delta}, \vec{\beta}, \vec{\omega}, \vec{\nu}, \dvec{z}, \ddvec{z}, \vec{z}^{(1:T)}}~\sum_{t=1}^{T} \Big\| \vec{A}^{(t)} \big( \underbrace{\overbrace{\vec{B}\vec{v}^{(t)}}^{\vec{c}^{(t)}} \circ \overbrace{\vec{D}\vec{W}^{(t)}}^{\vec{\phi}^{(t)}}}_{\Tvec{x}^{(t)}} \big) - \vec{y}^{(t)} \Big\|_2^2 + \textb{\lambda}\mathcal{R}\big(\vec{\phi}^{(1:T)}\big),
    \label{eq:optim}
\end{equation}

Here, $\vec{W}^{(t)}\leftarrow \mathcal{G}_{\vec{\omega}}(\vec{z}^{(t)})$, $\vec{v}^{(t)}\leftarrow \mathcal{G}_{\vec{\nu}}(\vec{z}^{(t)})$, $\vec{d}^{(1:L_1)}\leftarrow\mathcal{G}_{\vec{\delta}} (\dvec{z})$, and $\vec{b}^{(1:L_2)}\leftarrow\mathcal{G}_{\vec{\beta}}(\ddvec{z})$ represent the outputs of four generators. As mentioned previously, the matrix $\vec{D}$ is generated by concatenating the elements of $\vec{d}^{(1:L_1)}$ as columns, and the matrix $\vec{B}$ is generated by concatenating the elements of $\vec{b}^{(1:L_2)}$ as columns. The term $\mathcal{R}(\vec{\phi}^{(1:T)})$ represents the spatial and/or temporal regularization applied to the deformation fields. \textb{The strength of the regularization is controlled by $\lambda \geq 0$.} The ``$\circ$'' operation represents spatial warping of the composite image. Note, the warping operation is typically realized using 3D deformation fields and 3D images and not their vectorized representations. However, for notational simplicity, we express this operation between two vectors, $\vec{B}\vec{v}^{(t)}$ and $\vec{D}\vec{W}^{(t)}$.

After the network is trained, the $t^{\text{th}}$ 3D frame can be recovered by passing the optimized code vectors, $\Hat{\dvec{z}}$, $\Hat{\ddvec{z}}$, and $\Hvec{z}^{(t)}$, through trained sub-networks, $\mathcal{G}_{\Hvec{\delta}}$, $\mathcal{G}_{\Hvec{\beta}}$, $\mathcal{G}_{\Hvec{\omega}}$, and $\mathcal{G}_{\Hvec{\nu}}$. This on-demand production of one or more frames obviates the need to generate or save the entire image series, which may have thousands of 3D frames for the cine acquisition performed over several minutes.


\subsubsection{Implementation details of ML-DIP}
As shown in \figref{overview}, ML-DIP consists of four \textb{sub-networks} and three sets of code vectors. 
The architectures of the sub-networks are reported in \tabref{architecture}. 
The input to ConvDecoder, \dvec{z}, had $h=2$ channels, with each channel consisting of a real-valued 3D array. The size of the 3D array was determined by the number of upsampling steps between the input and output of ConvDecoder. Likewise, the input to U-Net, \ddvec{z}, had $h=2$ channels, with each channel consisting of a real-valued 3D array. The size of the 3D array was identical to the size of the target image. The entries of $\dvec{z}$ and $\ddvec{z}$ were initialized independently from a uniform distribution. \textb{The sizes of the real-valued deformation basis ($L_1$) and complex-valued image basis ($L_2$) were set at 32 and 4, respectively.} The entries of $\vec{z}^{(1:T)}$ were initialized from the six principal components of the self-gating signal extracted from the repeated sampling of a central k-space line \cite{chen2024cardiac}. To regularize the deformation fields, $\mathcal{R}(\vec{\phi}^{(1:T)})$ in \eqref{optim} was chosen to be the sum of squares of the finite differences computed along the three spatial directions, \textb{with $\lambda = 0.05$.} The sub-networks and the code vectors were jointly trained for 48,000 iterations with the Adam optimizer. The batch size, which corresponds to the number of contiguous frames used in each update step, was set to 20. A cosine annealing learning rate schedule \cite{loshchilov2017sgdr} was used to reduce the learning rate from its initial value of $1\times10^{-3}$ to the final value of $2\times10^{-4}$. The total number of learnable parameters was approximately 17 million. After training, the final network parameters and code vectors were saved. For inference, optimized $\Hat{\dvec{z}}$, $\Hat{\ddvec{z}}$, and $\Hvec{z}^{(\tau_1:\tau_2)}$ were passed through the trained network to generate $\Hvec{x}^{(\tau_1:\tau_2)}$, i.e., frames in time interval $\tau_1 \leq t \leq \tau_2$ for user-defined values of $\tau_1$ and $\tau_2$, such that $1\leq \tau_1 \leq \tau_2 \leq T$.

\subsection{Experiments}
To evaluate ML-DIP, we analyzed data from a 3D MRXCAT phantom \cite{wissmann2014mrxcat}, ten healthy subjects at rest (two also scanned during in-magnet exercise), and \textb{12 patients with a history of premature ventricular contractions (PVCs).} All in vivo studies were approved by the Institutional Review Board, and written informed consent was obtained. The subject characteristics of the participants are summarized in \tabref{human}.

\subsubsection{MRXCAT phantom}
MRXCAT is a numerical simulation framework for cardiac MR imaging based on the extended cardiac torso (XCAT) phantom \cite{wissmann2014mrxcat}. It provides realistic anatomical models of the heart and thorax and supports user-defined cardiopulmonary motion patterns. Additionally, MR-specific parameters such as tissue properties, coil configurations, and noise levels can be customized in MATLAB (MathWorks, Natick, MA). 

To evaluate the performance of ML-DIP, we simulated a 3D cine MRXCAT phantom with an isotropic spatial resolution of 2 mm and 358 frames spanning 5 distinct respiratory cycles and 20 cardiac beats. One PVC beat was included in each respiratory cycle by shortening and altering the cardiac cycle. The 3D cine was then repeated 25 times along the temporal dimension to simulate a prolonged five-minute scan consisting of $T_0 = 8,\!950$ frames. To reduce computation time, the imaging volume was cropped to $110\times 112\times 92$ voxels along the \textb{superior-inferior (SI),} anterior-posterior, and left-right directions, respectively. Complex multicoil k-space data were generated using 8 receive coils and undersampled using ordered pseudo-radial sampling (OPRA) \cite{joshi2022technical}. The SI direction was used as the frequency encoding direction ($k_x$), anterior-posterior as the phase encoding direction ($k_y$), and left–right as the slice encoding direction ($k_z$). To mimic the MRI acquisition, no undersampling was applied along the $k_x$ direction.  
A total of 11 readouts were simulated for each 3D frame, resulting in the net acceleration rate of $R=936$. \figref{sampling} shows the sampling pattern used for the phantom. The sixth readout in each frame was collected along the SI orientation at $k_y=k_z=0$. These central lines were subjected to bandpass filtering followed by principal component analysis (PCA) to extract six motion components, including two for respiratory and four for cardiac \cite{chen2024cardiac}. These extracted motion components were used to initialize $\vec{z}^{(1:T)}$ in ML-DIP training.

To investigate the impact of $T$ (``scan time'') on the performance of ML-DIP, the model was trained separately for eight different values of $T$, i.e., $T = T_0 = 8,\!950$, $T=\frac{4}{5}T_0 = 7,\!160$, $T=\frac{3}{5}T_0 = 5,\!370$, $T=\frac{2}{5}T_0 = 3,\!580$, $T=\frac{1}{5}T_0 = 1,\!790$, $T=\frac{1}{10}T_0 = 895$, $T=\frac{1}{25}T_0 = 358$, and $T\approx \frac{1}{90}T_0 = 100$. This was achieved by truncating the original image series before training.

\subsubsection{Imaging healthy subjects with ferumoxytol}
\begin{figure*}
    \centering
    \includegraphics[width=0.95\linewidth]{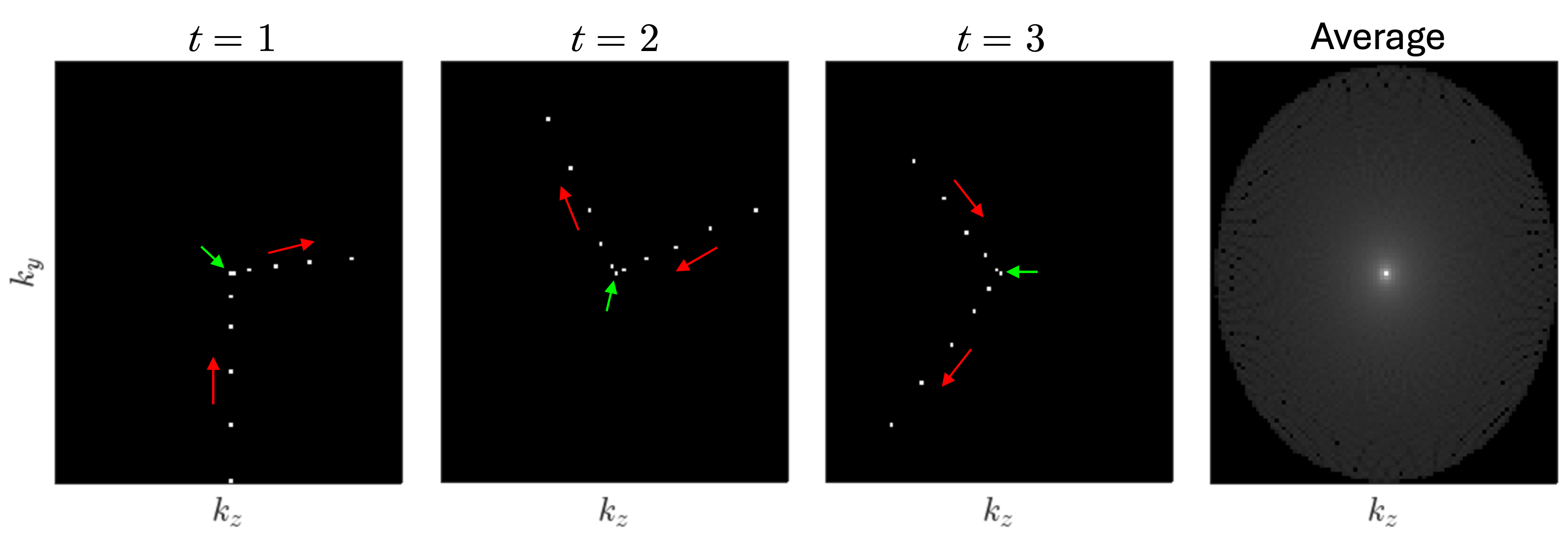}
    \caption{The Cartesian sampling pattern, OPRA, on a $112\times 92$ grid. Here, $t=1$, $t=2$, and $t=3$ represent the first three frames, each with eleven readouts (white dots), and the ``Average'' represents the average of all $T$ frames. The readout dimension $k_x$ is fully sampled and not shown. The red arrows show the acquisition order, highlighting smooth transitions from one readout to the next, even across frames. The green arrow points to the self-gating readouts at $k_y=k_z=0$. 
    }
    \label{fig:sampling}
\end{figure*}

ML-DIP was also validated using prospectively undersampled 3D real-time cine data collected from ten healthy volunteers, two of whom were additionally scanned during in-magnet exercise at a workload of 40\,W. \textb{The subjects were recruited to participate in an unrelated exercise imaging study \cite{chandrasekaran2025accelerated}, with the 3D cine acquisition added as an ancillary scan.} All volunteers were imaged with an ungated spoiled gradient echo-based 3D cine sequence on a 3\,T scanner (MAGNETOM Vida, Siemens Healthineers, Forchheim, Germany), following ferumoxytol infusion at 4 mg/kg. The 3D cine data were acquired under free-breathing conditions for five minutes with 11 OPRA readouts per 3D frame, as shown in \figref{sampling}. The imaging volume was a sagittal slab covering the entire heart, with frequency encoding in the \textb{SI direction,} phase encoding in the anterior-posterior direction, and slice encoding in the left-right direction. Other imaging parameters included: TE/TR of 1.2/3.1\,ms; temporal resolution of 32–34\,ms; acceleration rate of $R=1,\!047$; matrix size of $190\times 144\times 80$; spatial resolution of $1.4$–$2.2$\,mm along frequency encoding, $1.6$–$2.2$ mm along phase encoding, and $1.8$–$2.3$\,mm along slice encoding; and flip angle of 15$^\circ$. Each dataset contained approximately $9,\!000$ frames.

For comparison, each volunteer also underwent a free-breathing scan with a spoiled gradient-echo 2D real-time cine research sequence \cite{chandrasekaran2025accelerated}. A prospectively undersampled short-axis stack covering the whole heart and one or more long-axis views was collected using variable-density golden-ratio offset sampling ($R = 12$) \cite{joshi2022technical}. Data were reconstructed inline with Gadgetron \cite{hansen2013gadgetron} using a parameter-free compressed-sensing method \cite{chen2019sparsity}. The 2D scans had a spatial resolution 2.0–2.1\,mm, a temporal resolution 37–39\,ms, and a scan time of 6\,s per slice. The 2D and 3D acquisitions were completed within five minutes of each other.

\subsubsection{Imaging PVC patients \textb{post-gadolinium}}
To assess the ability of ML-DIP to recover irregular beats, the 3D cine acquisition was appended to the clinical exam of \textb{12 patients with a history of frequent PVCs. Six patients exhibited irregular, frequent PVCs during the scan, three were predominantly in bigeminy, and three did not exhibit frequent PVCs during the five-minute acquisition.} The patients were scanned on a 1.5\,T scanner (MAGNETOM Sola, Siemens Healthineers, Forchheim, Germany). Data were acquired under free-breathing conditions for five minutes using a balanced steady-state free-precession sequence with 11 OPRA readouts per 3D frame. The imaging volume was a sagittal slab covering the entire heart, with frequency encoding in the \textb{SI direction,} phase encoding in the anterior-posterior direction, and slice encoding in the left-right direction. Imaging parameters included: TE/TR = 1.3/3.1\,ms, temporal resolution = 32–34\,ms, acceleration rate = $R=1,\!047$, matrix size = $190 \times 144 \times 80$, spatial resolution = 1.4–1.8\,mm (frequency), 1.5–1.8\,mm (phase), and 2.0\,mm (slice), and flip angle = 34–40$^\circ$. Each dataset contained approximately $9\!,000$ frames. For comparison, 2D real-time imaging was also performed using a balanced steady-state free-precession sequence with 3\,s per slice. The spatial and temporal resolutions of the 2D scans were similar to those used in the healthy subject study. Unlike the previous study, ferumoxytol was not used in this patient study, and 3D acquisition occurred approximately 20–30 minutes after administration of a gadolinium-based contrast agent (Gadovist, Bayer AG, Berlin, Germany). 


\subsection{Data processing and analysis}
\subsubsection{\textb{Preprocessing}}
To reduce computation time for in vivo studies, the k-space data in the $k_x$–$k_y$–$k_z$ domain were first transformed to the spatial domain along the readout ($k_x$) dimension using a 1D fast Fourier transform (FFT). The data were then cropped along this dimension and transformed back to the $k_x$–$k_y$–$k_z$ domain using an inverse 1D FFT. This step was performed by presenting users with a time-averaged sagittal image and prompting them to select two points, one above and one below the heart. Although not required, this subject-specific cropping reduced computation time by limiting the readout dimension. To further accelerate computation, the physical coils were compressed to fifteen virtual coils using PCA \cite{buehrer2007array} and then to eight using region-optimized virtual coils \cite{kim2021region}. Since region-optimized coil compression has the tendency to select virtual coils that are mostly noise but have a favorable signal-to-interference ratio, PCA-based compression was performed first as a denoising step. Coil sensitivity maps of the eight virtual coils were then estimated from the time-averaged k-space using ESPIRiT \cite{uecker2014espirit}.

\subsubsection{Image reconstruction}
For each dataset, training was performed on a single RTX 6000 Ada (Nvidia, Santa Clara, California). Depending on the size of the imaging matrix, training time was approximately 7 to 10 hours. For the MRXCAT phantom, 100 consecutive frames were generated post-training. The inference time to generate these frames was approximately 2-3 seconds. For the data from healthy subjects and patients, 200 consecutive frames were generated post-training, with an inference time of 5–8 seconds. For comparison, the same 3D cine data were binned into 20 cardiac and 4 respiratory phases and reconstructed using compressed sensing \cite{feng20185d}, with regularization applied along the spatial, respiratory, and cardiac dimensions. We refer to this method as 5D-Cine. The reconstruction time for 5D-Cine was approximately two hours on the same workstation. To improve the feasibility of 5D-Cine in PVC patients, arrhythmia rejection was applied by discarding k-space data from beats that deviated significantly from the average beat length \cite{pruitt2021fully}. \textb{Only the expiratory phase was analyzed from 5D-Cine, as it provided the highest image quality among the four respiratory phases.}

\subsubsection{Image analysis}
For the phantom data, where the noiseless ground truth was available, image quality of the generated frames was assessed using peak signal-to-noise ratio (PSNR) and structural similarity index measure (SSIM) for all eight values of $T$. For the data collected from human subjects, both ML-DIP and 5D-Cine results were interpolated into standard short-axis and long-axis views using slice position information from the 2D real-time cine data. To evaluate the image quality of 5D-Cine, 2D real-time cine, and ML-DIP, one short-axis and one long-axis view from each subject were blindly scored by two experienced Level-3 CMR-trained cardiologists. The three cine series were presented as movies on a single slide, with the order of 5D-Cine, 2D real-time cine, and ML-DIP randomized. Each reader assigned a score from 1 (worst) to 5 (best) to each cine based on overall image quality: 1 – Non-diagnostic, 2 – Poor, 3 – Fair, 4 – Good, 5 – Excellent. The interpolated short-axis stacks from 5D-Cine and ML-DIP were converted to digital imaging and communications in medicine (DICOM) format and analyzed in suiteHEART (NeoSoft, Pewaukee, Wisconsin) along with the 2D real-time stack. Left-ventricular end-diastolic volume (EDV), end-systolic volume (ESV), stroke volume (SV), and ejection fraction (EF) were computed from 2D real-time cine, 5D-Cine, and ML-DIP using the short-axis stacks. Quantification was performed only for subjects whose average image quality score across both readers and both views was at least 3.




\section{Results}
\tabref{psnr} shows image quality metrics from the MRXCAT phantom for different values of $T$. As expected, image quality decreased with shorter scan durations due to reduced k-space coverage. However, there was no significant drop in PSNR or SSIM when $T$ was reduced from 8,950 (five minutes) to 3,580 (two minutes), and only a modest drop in the metrics at $T=1,\!790$ (one minute). A more substantial decline in image quality metrics was observed when $T$ was at or below 895 (30 seconds). \figref{mrxcat} shows representative images at different values of $T$. These images show that ML-DIP effectively recovered both respiratory and cardiac motion, including PVC beats. A PVC beat, marked by yellow arrows, is evident in the space-time (x-t) profiles. Only the images at $T=358$ and $T=100$ showed noticeable artifacts and noise amplification. These results suggest that ML-DIP can be effective at shorter acquisition times. A movie, Video S1, corresponding to \figref{mrxcat}, is provided in the Supplementary material.

\begin{figure*}
    \centering
    \includegraphics[width=0.95\linewidth]{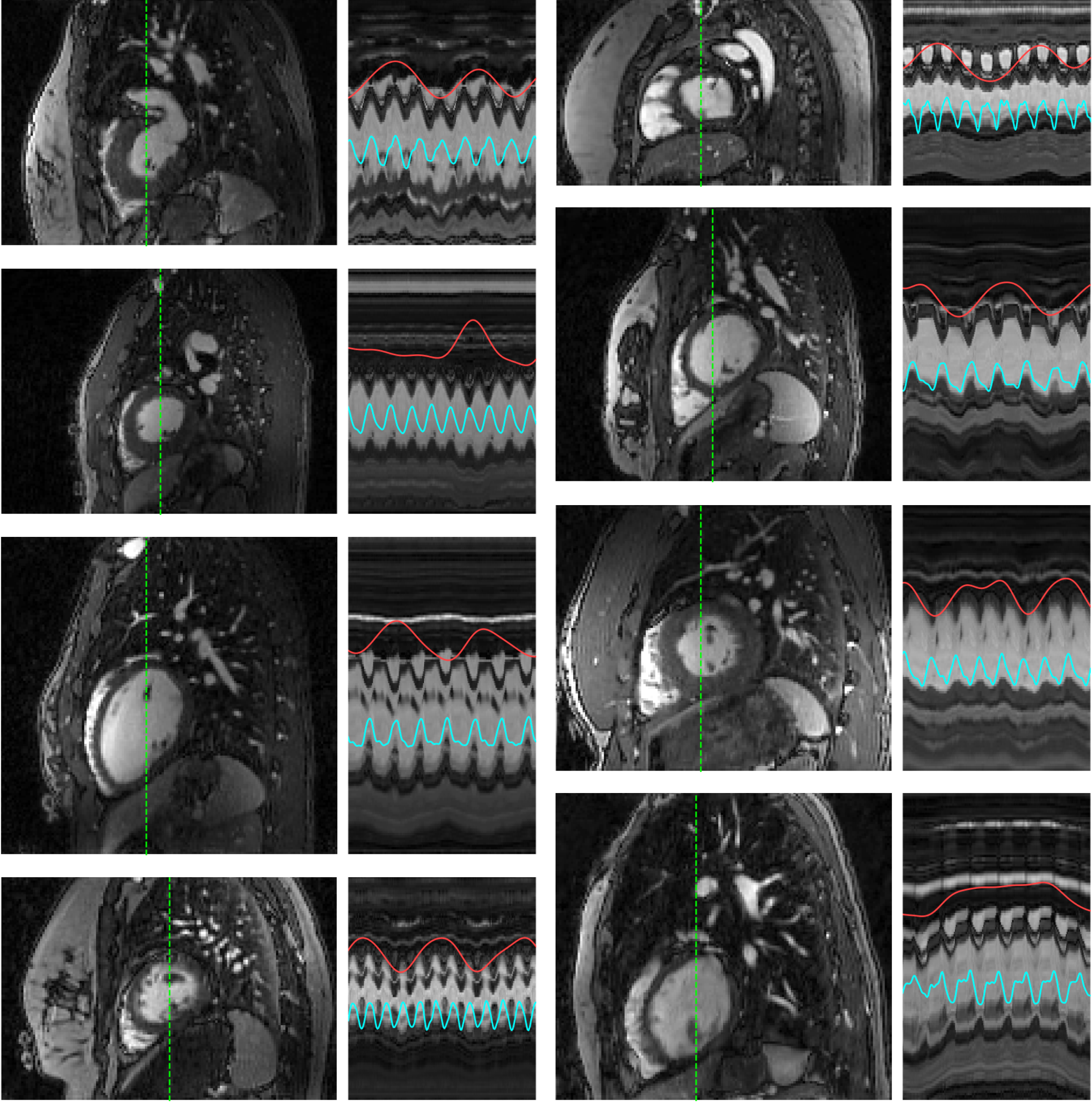}
    \caption{Representative ML-DIP images in sagittal orientation from eight subjects imaged at rest. Time profiles along the dashed green lines are shown to the right of the images. Each x–t profile spans 200 frames (6.6\,s). The red and cyan curves represent self-gating-based respiratory and cardiac signals, respectively. 
    }
    \label{fig:rest_sag}
\end{figure*}

\begin{figure*}
    \centering
    \includegraphics[width=0.95\linewidth]{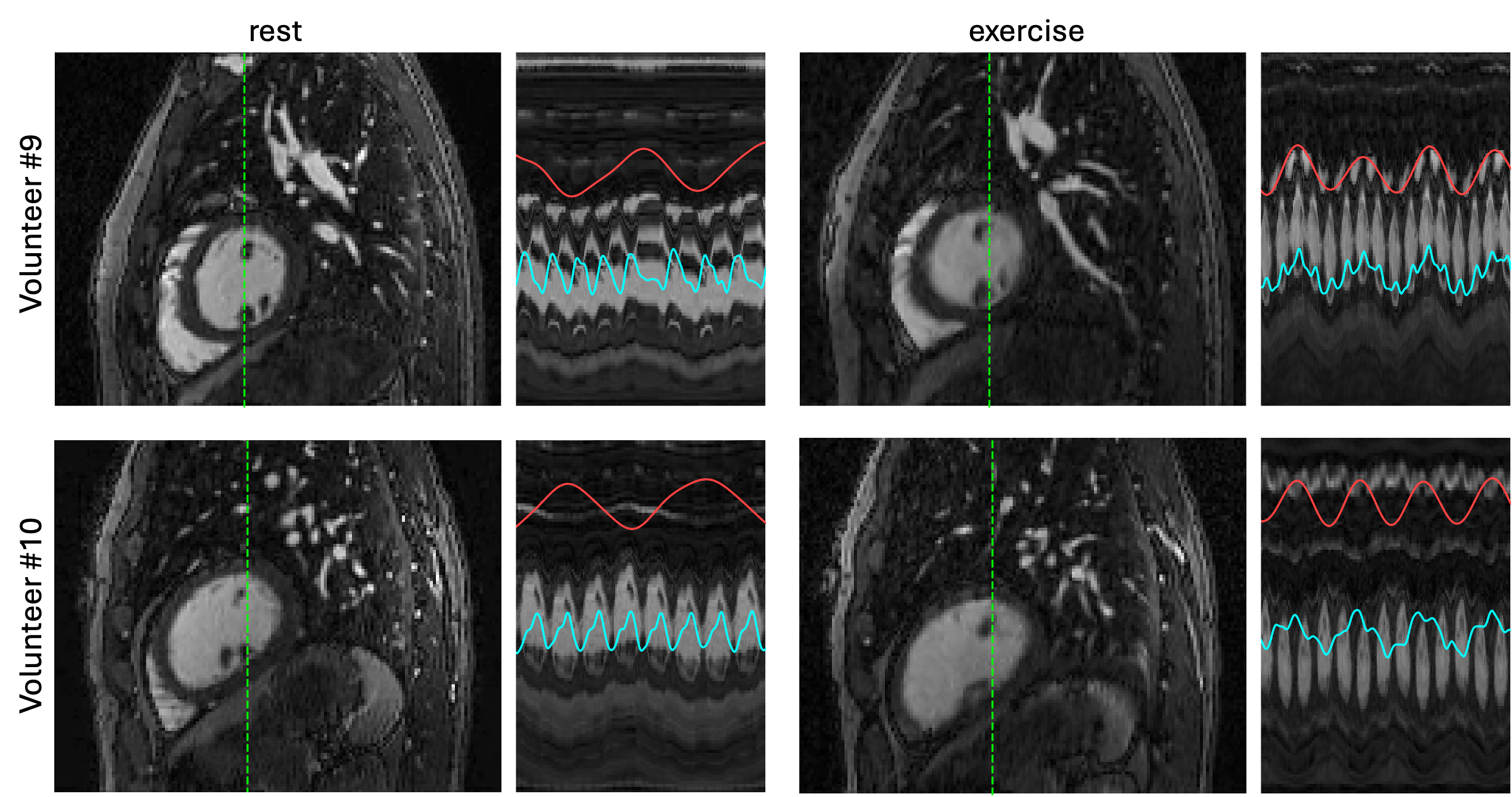}
    \caption{Representative ML-DIP images in sagittal orientation from two subjects imaged at rest and during in-magnet exercise. Time profiles along the dashed green lines are shown to the right of the images. Each x–t profile spans 200 frames (6.6\,s). The red and cyan curves represent self-gating-based respiratory and cardiac signals, respectively. Significantly faster heart rates are observed during exercise. 
    }
    \label{fig:exercise_sag}
\end{figure*}

\figref{fixed} highlights the advantage of using a frame-specific composite image in ML-DIP. Relying on a single fixed template resulted in visible image distortions, as highlighted by the yellow arrows. Because the fixed-template approach performed poorly compared to ML-DIP, it was excluded from further comparisons. \figref{rest_sag} shows representative ML-DIP reconstructions from eight healthy volunteers scanned at rest. The corresponding x-t profiles highlight that ML-DIP preserves both cardiac and respiratory motions, closely following the self-gating signals. \figref{exercise_sag} shows results from two additional volunteers scanned both at rest and during in-magnet exercise. During exercise, faster heart rates and exaggerated breathing patterns were captured without visible degradation in image quality. \figref{pvc_sag} shows representative results from \textb{six of the 12 PVC patients.} Despite lower blood–myocardium contrast, ML-DIP successfully captured beat-to-beat variations, including the timing and morphology of PVCs. PVC beats were easily identified on the reconstructed x–t profiles and corroborated by the self-gating signals.  

In healthy subjects, blind scoring by cardiologists consistently yielded higher scores for ML-DIP, as summarized in \tabref{scores}. The advantage of ML-DIP over 5D-Cine was more pronounced during exercise, where 5D-Cine exhibited visible motion artifacts due to unaccounted torso movement. In PVC patients, where ferumoxytol was not used, 2D real-time scored higher than ML-DIP, which can be attributed to lower contrast in 3D imaging from blood pool saturation \cite{nezafat2008inflow}. \textb{Nonetheless, ML-DIP received an average score of 4.50, with the lowest score of 3.75 for PVC \#3. Despite arrhythmia rejection, 5D-Cine performed poorly in PVC patients, with eight subjects scoring below 3.00 and four scoring below 2.00. Only four PVC patients---three without arrhythmia during acquisition and one in stable bigeminy---received a score of 3.00 or higher.} \figref{comparison} shows some of the short-axis images scored by expert readers. In healthy subjects, ML-DIP produced sharp, motion-artifact-free images, even during exercise. In PVC patients, ML-DIP reconstructed PVC beats, including the compensatory pauses following PVCs in one subject and alternating premature beats consistent with bigeminy in another. \textb{Beat-to-beat variations observed with ML-DIP were also consistent with the ECG traces, which are superimposed on the temporal profiles of the two PVC subjects in \figref{comparison}.} In contrast, 5D-Cine not only removed beat-to-beat variations but also exhibited extensive motion artifacts and blurring due to inconsistent cardiac binning. A movie, Video S2, corresponding to \figref{comparison}, is shown in the Supplementary material. \textb{To highlight the improvement in blood-myocardium contrast from ferumoxytol, Video S3 in the Supplementary material presents two ML-DIP cine series: one from a healthy subject scanned with ferumoxytol at 3T and the other from a PVC patient scanned post-gadolinium at 1.5T.}

\begin{figure*}[t]
    \centering
    \begin{minipage}[t]{0.45\textwidth}
        \centering
        \begin{overpic}[width=\linewidth]{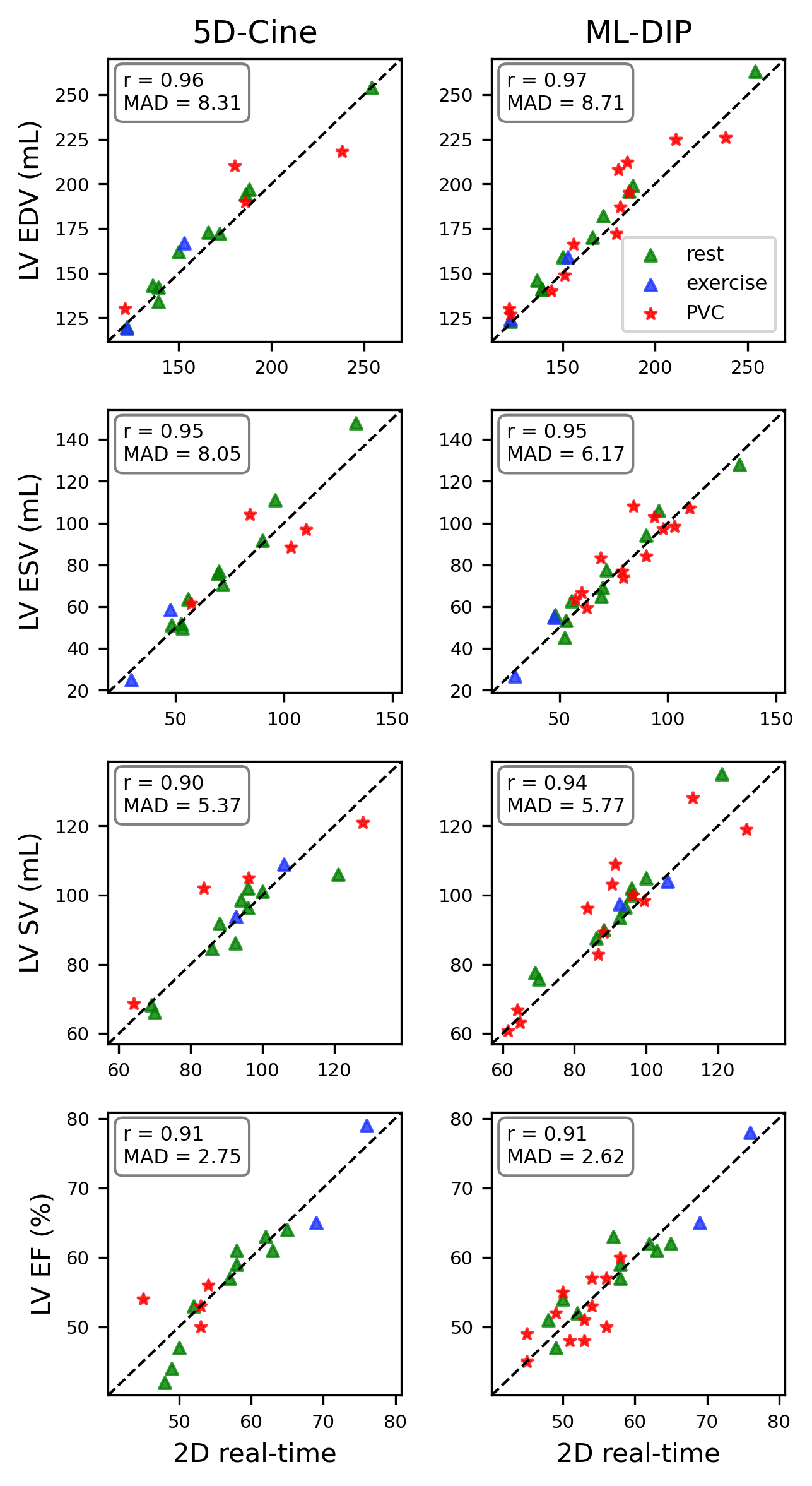}
            \put(0,97){\textsf{(a)}} 
        \end{overpic}
    \end{minipage}
    \hspace{0.02\textwidth}
    \begin{minipage}[t]{0.45\textwidth}
        \centering
        \begin{overpic}[width=\linewidth]{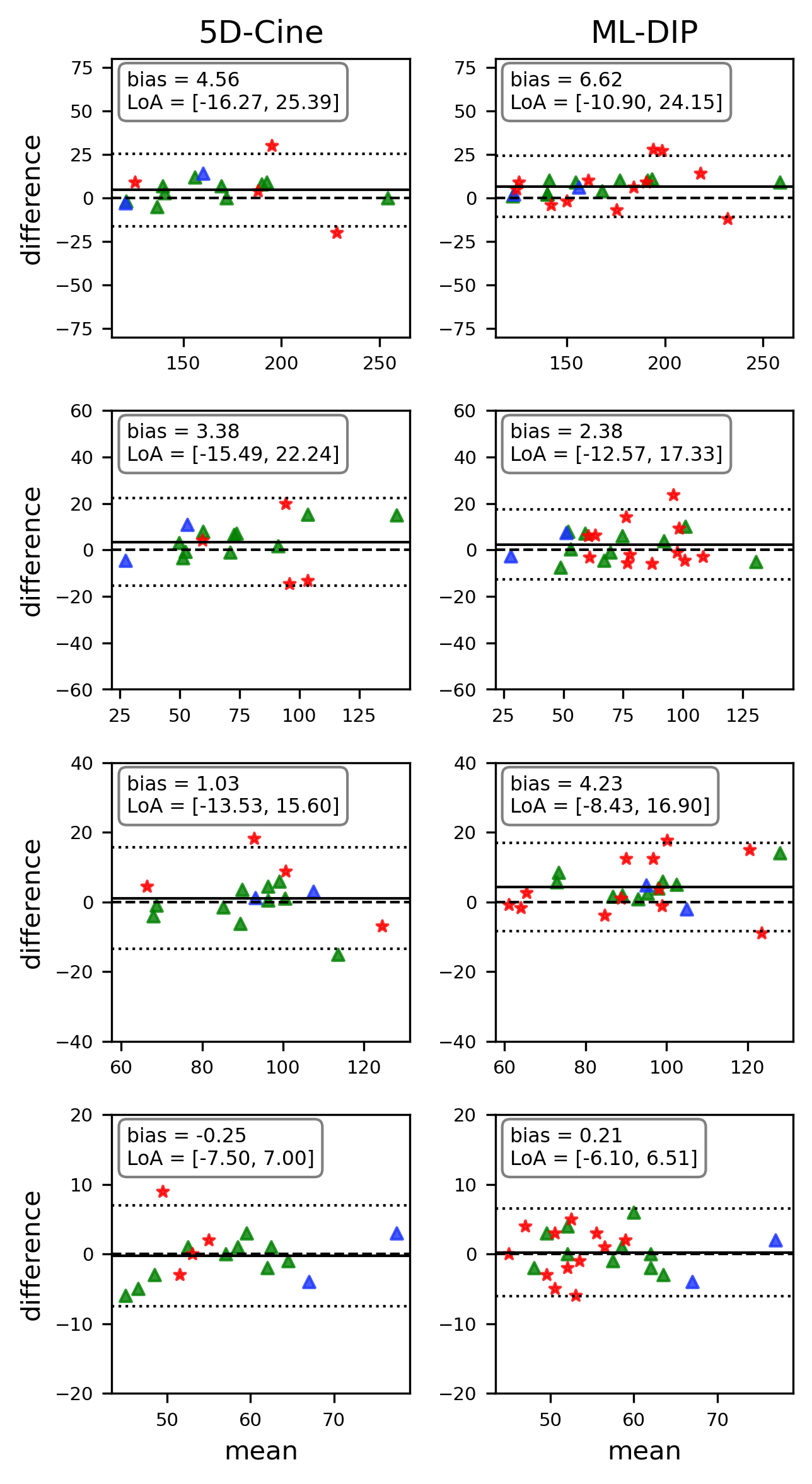}
            \put(0,97){\textsf{(b)}}
        \end{overpic}
    \end{minipage}

    \caption{Left ventricular (LV) quantification from 5D-Cine and ML-DIP, with 2D real-time used as the reference.
    \textb{In the case of 5D-Cine, only four (out of twelve) PVC patients are included because the image quality from eight other patients was not adequate for analysis.  (a) Correlation plot. (b) Bland–Altman plot, where ``difference'' represents volumetric reconstruction\,-\,2D.} 
    }
    \label{fig:quant}
\end{figure*}

\begin{figure*}
    \centering
    \includegraphics[width=0.8\linewidth]{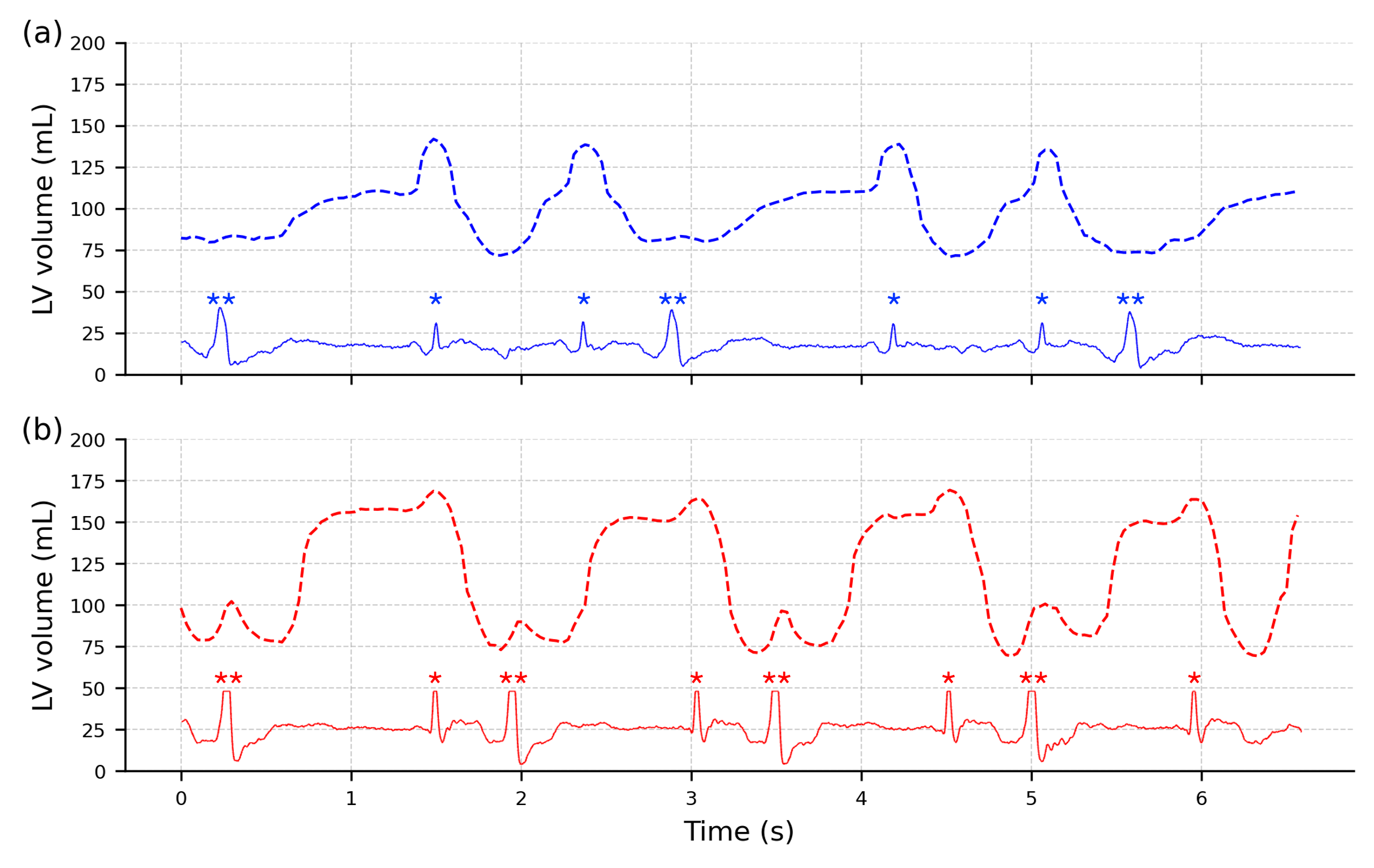}
    \caption{\textb{LV volumes (dashed lines) across 200 ML-DIP frames from two different PVC patients in (a) and (b). The corresponding ECG traces (solid lines), expressed in arbitrary units, are also shown at the bottom. R-waves corresponding to sinus beats are marked with `*,' and those corresponding to PVCs are marked with `**.' 
    }
    }
    \label{fig:beat-to-beat}
\end{figure*}

The cardiac function quantification results are summarized in \figref{quant}. In healthy subjects, LV functional measurements from both 5D-Cine and ML-DIP showed excellent agreement with those from 2D real-time cine, including during exercise.
In PVC patients, functional analysis was feasible for all \textb{12 subjects using ML-DIP, but only for four subjects using 5D-Cine due to poor image quality.} \textb{Across all 24 acquisitions (healthy + patients), the mean absolute difference between ML-DIP and 2D real-time was 8.7 mL for EDV, 6.2 mL for ESV, 5.8 mL for SV, and 2.6\% for EF. Pearson correlation coefficients between ML-DIP and 2D real-time were $>0.9$ for all metrics. All volumetric measurements were limited to the first dominant sinus contraction, identified as a large drop in LV volume. In PVC patients, beat-to-beat or arrhythmic-beat quantification was not feasible with 2D real-time imaging because that would require substantially longer scan times per slice and manual alignment of matching beats across slices \cite{contijoch2016quantification}. In contrast, ML-DIP yields temporally coherent 3D volumes, enabling visualization of beat-to-beat variations in cardiac function. \figref{beat-to-beat} shows representative examples demonstrating these variations in two PVC patients together with the corresponding ECG traces.}

\section{Discussion}
Three-dimensional real-time imaging has the potential to offer a paradigm shift for CMR imaging. However, due to very high acceleration rates ($R>1,\!000$), 3D real-time imaging has not been previously possible. In this work, we develop a scan-specific learning-based framework, called ML-DIP, that can facilitate 3D real-time imaging from a free-breathing ungated scan. A key innovative feature of ML-DIP is its ability to model multiple, disparate image dynamics using scalable, low-rank representations.

Our phantom results demonstrate that ML-DIP can recover image series with high values of PSNR and SSIM. As evident from \tabref{psnr}, image quality degrades as the value of $T$ gets smaller, with the images from $T=358$ and $T=100$ datasets showing significant noise amplification. This is expected because the learning in ML-DIP is based on the entire time series and not the individual frames. Since each frame contributes a unique, complementary sampling pattern, smaller values of $T$ leave most indices in k-space unsampled for any one motion state. Nonetheless, the PSNR and SSIM values of ML-DIP stay within 0.5 dB and 0.03, respectively, of the best values until $T = 3,\!580$, which corresponds to a two-minute scan. Although we have performed all subsequent in vivo studies with a fixed acquisition time of five minutes, this phantom study suggests that there is a margin to reduce the acquisition time to well below five minutes.


Our healthy subject study demonstrates that ML-DIP can generate high-quality images while preserving cycle-to-cycle variations in respiratory and cardiac motion. For images collected at rest, all three methods received a high image quality score of 4.00 or higher. However, images from ML-DIP were visibly sharper than those from 5D-Cine and even 2D cine, where the former suffers from blurring due to intra-bin motion and the latter from inhomogeneous blood pool intensity due to the mixing of excited and unexcited blood in the presence of ferumoxytol. The quality gap between ML-DIP and 5D-Cine was more pronounced during exercise. In this setting, ML-DIP captured cardiac, respiratory, and bulk motion due to pedaling on an ergometer in real time, whereas 5D-Cine reconstructed only a representative cardiorespiratory cycle, with unaccounted bulk motion manifesting as artifacts. Nonetheless, due to high blood–myocardium contrast, 5D-Cine received scores of at least 3.00 in all cases. 

\begin{figure*}
    \centering
    \includegraphics[width=0.95\linewidth]{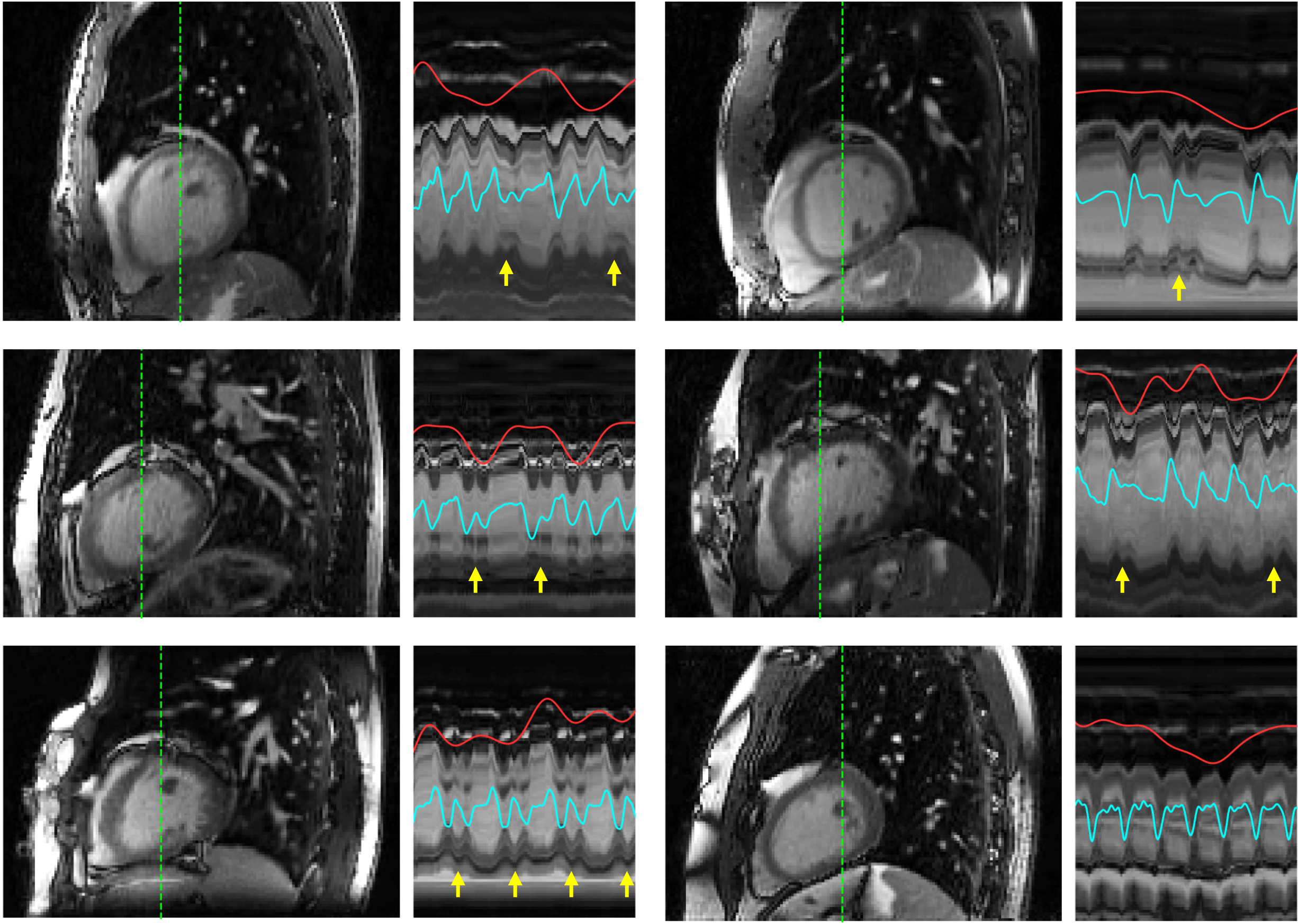}
    \caption{\textb{Representative ML-DIP images in sagittal orientation from six out of twelve PVC patients.} Space-time (x-t) profiles along the dashed green lines are shown to the right of the images. Each x–t profile spans 200 frames (6.6\,s). The red and cyan curves represent self-gating-based respiratory and cardiac signals, respectively. PVC beats are indicated by yellow arrows. The compensatory pause is also visible after some of the PVC beats. \textb{One of the patients (bottom-left) was in bigeminy, evident in the x–t profile, and another patient (bottom-right) did not experience PVCs during the scan.} 
   }
    \label{fig:pvc_sag}
\end{figure*}

The 3D imaging of PVC subjects is more challenging due to lower blood–myocardium contrast without ferumoxytol, reduced signal-to-noise ratio at lower field strength, and irregular heart rhythms. Despite these challenges, ML-DIP successfully captured both respiratory and cardiac dynamics, including PVC beats. Although the self-gating signal is less precise in subjects with irregular breathing or arrhythmias, it remains sensitive to detecting irregular events. We therefore used self-gating \textb{as well as synchronously acquired ECG as surrogates} to validate the timing and consistency of the reconstructed PVC dynamics. \textb{We observed that the PVC occurrences in the ML-DIP reconstructions aligned temporally with both the self-gating cardiac signal (\figref{pvc_sag}) and the ECG signal (\figref{comparison}), and their temporal patterns matched those observed in 2D real-time images.} However, 2D imaging captures arrhythmias in only a limited number of slices and may miss events occurring outside those planes or acquisition windows. ML-DIP overcomes this limitation by providing volumetric coverage with consistent temporal sampling, enabling whole-heart assessment of arrhythmias from a single scan. Furthermore, because ML-DIP does not rely on binning or arrhythmia rejection, it preserves true beat-to-beat variation and avoids the artifacts and information loss associated with retrospective gating. While larger validation studies are needed, this preliminary evidence suggests that ML-DIP can reliably recover complex, subject-specific motion patterns in challenging patient populations.

\section{Limitations}
The current study has several limitations. \textb{First, our sample size is small and includes only one type of arrhythmia, i.e., PVC. Second, completely non-contrast acquisitions were not considered and are expected to offer lower blood-myocardium contrast. Third,} the reconstruction time for ML-DIP is long, which limits translational potential in the short term. However, one might be able to accelerate ML-DIP by pre-training the ConvDecoder and/or U-Net on the coarse results obtained from 5D-Cine before jointly training all four sub-networks in an end-to-end fashion. Another avenue to accelerate ML-DIP is to adopt a hierarchical approach where the spatial and temporal resolutions of the images are progressively increased over iterations. Fourth, additional validation is needed to confirm that ML-DIP preserves strain patterns and regional wall motion abnormalities, which are important for detecting subtle functional impairments.  \textb{Fifth, all the presented studies were conducted at 32-34 ms temporal resolution; the impact of changing temporal resolution on the quality of ML-DIP reconstructions needs to be further explored. Sixth,}  the current architectures used in ML-DIP and the values of hyperparameters are not fully optimized. This is primarily due to the long reconstruction times. Future studies of ML-DIP will include a larger sample size and a more diverse set of CMR applications.

\section{Conclusions}
We have presented and evaluated a scan-specific framework, called ML-DIP, for 3D real-time CMR. A key feature of ML-DIP is its ability to model both motion and contrast changes. The low-rank modeling for both the image and deformation representations makes the network architecture simpler and easier to train. Phantom and in vivo results demonstrate the potential of ML-DIP to preserve real-time dynamics from highly undersampled 3D data.

\newpage

\section*{Acknowledgments and funding}
This work was funded by NIH grants R01-EB029957, R01-HL151697, and R01-HL148103.

\section*{Ethics declarations}
\begin{itemize}
\item \textbf{Competing interests}: The authors declare no competing interests.
\item \textbf{Ethics approval and consent}: For the human subject data, approval was granted by the Institutional Review Board (IRB) at The Ohio State University (2020H0402 and 2019H0076). Informed consent to participate in the study and publish results was obtained from all individual participants.
\item \textbf{Data and code availability}: For code, visit \url{https://github.com/OSU-CMR/ml-dip}
\item \textbf{Author contributions}: C. Chen implemented ML-DIP and generated initial results. M. Vornehm assisted with implementation and manuscript review. Z. Bu reconstructed the additional patient data and generated updated figures. M. Sultan provided feedback on improving and optimizing ML-DIP and reviewed the manuscript. M. Arshad assisted with data acquisition and curation. Y. Liu performed pulse sequence programming and scanned the subjects. P. Chandrasekaran assisted with volunteer recruitment, data acquisition, and image analysis. Y. Han contributed to experiment design and result interpretation. R. Ahmad supervised all aspects of the study.
\end{itemize}
\clearpage

\bibliographystyle{IEEEtran}
\bibliography{references}
\clearpage

\begin{table*}[h]
\small
\centering
\begin{tabular}{llll}
\multicolumn{4}{l}{$\mathcal{G}_{\vec{\delta}}$: \textbf{Deformation Basis Generator (ConvDecoder)}} \\
\toprule
\textbf{Block} & \textbf{Operation} & \textbf{In Ch.} & \textbf{Out Ch.} \\
\midrule
Decoder 1 & CBR × 3 + Upsample & 2 & 256 \\
Decoder 2 & CBR × 3 + Upsample & 256 & 256 \\
Decoder 3 & CBR × 3 + Upsample & 256 & 128 \\
Decoder 4 & CBR × 3 + Upsample & 128 & 128 \\
Decoder 5 & CBR × 3 + Upsample & 128 & 64 \\
Decoder 6 & CBR × 2 + Conv3D & 64 & $L_1$ \\
\bottomrule
\end{tabular}
\vspace{1em}

\begin{tabular}{llll}
\multicolumn{4}{l}{$\mathcal{G}_{\vec{\omega}}$: \textbf{Deformation Coefficient Generator}} \\
\toprule
\textbf{Layer} & \textbf{Operation} & \textbf{Input Dim} & \textbf{Output Dim} \\
\midrule
1 & FC + LeakyReLU & 6 & 32 \\
2 & FC + LeakyReLU & 32 & 64 \\
3 & FC + LeakyReLU & 64 & 128 \\
4 & FC + LeakyReLU & 128 & 256 \\
5 & FC + LeakyReLU & 256 & 256 \\
6 & FC + LeakyReLU & 256 & 128 \\
7 & FC & 128 &$3L_1$ \\
\bottomrule
\end{tabular}
\vspace{1em}

\begin{tabular}{llll}
\multicolumn{4}{l}{$\mathcal{G}_{\vec{\beta}}$: \textbf{Image Basis Generator (U-Net)}} \\
\toprule
\textbf{Block} & \textbf{Operation} & \textbf{In Ch.} & \textbf{Out Ch.} \\
\midrule
Encoder 1 & CBR × 1 & 2 & 16 \\
Encoder 2 & CBR × 1 + AvgPool & 16 & 32 \\
Encoder 3 & CBR × 1 + AvgPool & 32 & 64 \\
Encoder 4 & CBR × 1 + AvgPool & 64 & 64 \\
Encoder 5 & CBR × 1 + AvgPool & 64 & 64 \\
Bottleneck & CBR × 1 & 64 & 64 \\
Decoder 1 & Upsample + CBR × 1 & 64 + 64 & 64 \\
Decoder 2 & Upsample + CBR × 1 & 64 + 64 & 64 \\
Decoder 3 & Upsample + CBR × 1 & 64 + 64 & 32 \\
Decoder 4 & Upsample + CBR × 1 & 64 + 32 & 16 \\
Decoder 5 & CBR × 1 + Conv3D & 16 & $2L_2$ \\
\bottomrule
\end{tabular}
\vspace{1em}

\begin{tabular}{llll}
\multicolumn{4}{l}{$\mathcal{G}_{\vec{\nu}}$: \textbf{Image Coefficient Generator}} \\
\toprule
\textbf{Layer} & \textbf{Operation} & \textbf{Input Dim} & \textbf{Output Dim} \\
\midrule
1 & FC + LeakyReLU & 6 & 32 \\
2 & FC + LeakyReLU & 32 & 64 \\
3 & FC + LeakyReLU & 64 & 128 \\
4 & FC + LeakyReLU & 128 & 256 \\
5 & FC + LeakyReLU & 256 & 128 \\
6 & FC + LeakyReLU & 128 & 64 \\
7 & FC & 64 & $2L_2$ \\
\bottomrule
\end{tabular}
\caption{Summary of the four sub-networks used in ML-DIP. CBR = Conv3D + BatchNorm3D + LeakyReLU. FC = Fully connected. All Conv3D layers use kernel size $3\times3\times3$ with padding = 1. Upsampling and average pooling operations increase or decrease the size, along each dimension, by a factor of 2.}
\label{tab:architecture}
\end{table*}
\clearpage

\begin{table}[ht]
\centering
\begin{tabular}{|l|l|l|}
\hline
      & Healthy subjects & \textb{PVC patients} \\
\hline
\hline
Number & 10 & \textb{12} \\
\hline
Age (years) &  $31\pm10$ & \textb{$48\pm18$}  \\
\hline
Sex (M/F) & 4/6 & \textb{4/8} \\
\hline
BMI (kg/m$^2$) & $27.2\pm 5.3$ & \textb{$25.1\pm4.5$} \\
\hline
BSA (m$^2$) & $1.5\pm0.2$ & \textb{$1.4\pm 0.3$} \\

\hline
\end{tabular}
\caption{Human subjects characteristics.}
\label{tab:human}
\end{table}
\clearpage

\begin{table}[ht]
\centering
\begin{tabular}{|c|c|c|c|c|c|c|c|c|}
\hline
     & $T=8,\!950$ & $T=7,\!160$ & $T=5,\!370$ & $T=3,\!580$ & $T=1,\!790$ & $T=895$ & $T=358$ & $T=100$ \\
\hline
\hline
PSNR (dB) & 29.9 & 30.1 & 29.9 & 29.5 & 29.2 & 27.9 & 24.7 & 19.9 \\
\hline
SSIM     & 0.95 & 0.95 & 0.94 & 0.92 & 0.89 & 0.82 & 0.71 & 0.54 \\
\hline
\end{tabular}
\caption{PSNR and SSIM values for eight different numbers of total frames, $T$. These values of $T$, from left to right,  correspond to the scan times of $300$, $240$, $180$, $120$, $60$, $30$, $12$, and $3.4$ s, respectively.}
\label{tab:psnr}
\end{table}
\clearpage

\begin{table}[ht]
\centering
\begin{tabular}{|l|c|c|c|}
\hline
     Dataset & 5D-Cine & 2D real-time & ML-DIP \\
\hline
\hline
Vol. \#1 & 4.25 & 4.25 & \vec{4.75} \\
\hline
Vol. \#2 & 4.00 & 4.00 & \vec{5.00} \\
\hline
Vol. \#3 & \vec{5.00} & 4.25 & \vec{5.00} \\
\hline
Vol. \#4 & \vec{4.75} & 4.00 & 4.50 \\
\hline
Vol. \#5 & 4.00 & 4.25 & \vec{5.00} \\
\hline
Vol. \#6 & \vec{5.00} & 4.50 & \vec{5.00} \\
\hline
Vol. \#7 & 3.50 & 4.00 & \vec{5.00} \\
\hline
Vol. \#8 & \vec{4.75} & 4.25 & 4.50 \\
\hline
Vol. \#9 & \vec{5.00} & 4.50 & \vec{5.00} \\
\hline
Vol. \#10 & 4.50 & 4.25 & \vec{5.00} \\
\hline
\hline
Vol. Avg. & 4.48 & 4.23 & \vec{4.88}\\
\hline
\noalign{\vskip 8pt}
\hline
Vol. \#9-Ex. & 3.00 & 4.00 & \vec{4.25} \\
\hline
Vol. \#10-Ex. & 3.25 & 4.50 & \vec{5.00} \\
\hline
\hline
Vol.-Ex. Avg. & 3.13 & 4.25 & \vec{4.63}\\
\hline
\noalign{\vskip 8pt}
\hline
PVC \#1 & 1.00 & \vec{4.00} & \vec{4.00} \\
\hline
PVC \#2 & 3.75 & \vec{4.75} & 4.00 \\
\hline
PVC \#3 & 2.25 & \vec{4.50} & 3.75 \\
\hline
PVC \#4 & 2.00 & \vec{4.25} & \vec{4.25} \\
\hline
PVC \#5 & 3.00 & \vec{5.00} & 4.50 \\
\hline
\textb{PVC \#6} & \textb{2.75} & \textb{\vec{5.00}} & \textb{\vec{5.00}} \\
\hline
\textb{PVC \#7} & \textb{1.75} & \textb{\vec{4.75}} & \textb{\vec{4.75}} \\
\hline
\textb{PVC \#8} & \textb{4.00} & \textb{\vec{5.00}} & \textb{\vec{5.00}} \\
\hline
\textb{PVC \#9} & \textb{1.25} & \textb{\vec{5.00}} & \textb{4.50} \\
\hline
\textb{PVC \#10} & \textb{3.00} & \textb{4.00} & \textb{\vec{4.50}} \\
\hline
\textb{PVC \#11} & \textb{2.75} & \textb{\vec{4.75}} & \textb{4.50} \\
\hline
\textb{PVC \#12} & \textb{1.00} & \textb{\vec{5.00}} & \textb{4.75} \\
\hline
\hline
\textb{PVC Avg.} & \textb{2.38} & \textb{\vec{4.67}} & \textb{4.50}\\
\hline
\noalign{\vskip 8pt}
\hline
\textb{Overall Avg.} & \textb{3.31} & \textb{4.45} & \textb{\vec{4.67}}\\
\hline
\end{tabular}

\caption{Image quality scoring for the human subject studies. For an individual dataset, each number represents an average over two cardiac views and two expert readers. Here ``Vol.\ Avg.'' represents an average over 10 healthy volunteers imaged at rest, ``Vol.-Ex.\ Avg.'' represents an average over two healthy volunteers imaged during exercise, \textb{``PVC Avg.'' represents an average over 12 PVC patients, and ``Overall Avg.'' represents an average over all 24 cases.} In each row, the highest number is highlighted in bold font. \textb{PVC \#5, \#7, and \#11 were predominantly in bigeminy, and \#2, \#8, and \#10 did not exhibit frequent arrhythmias during the acquisition.}}
\label{tab:scores}
\end{table}
\clearpage

\clearpage

\clearpage

\begin{figure*}
    \centering
    \includegraphics[width=0.95\linewidth]{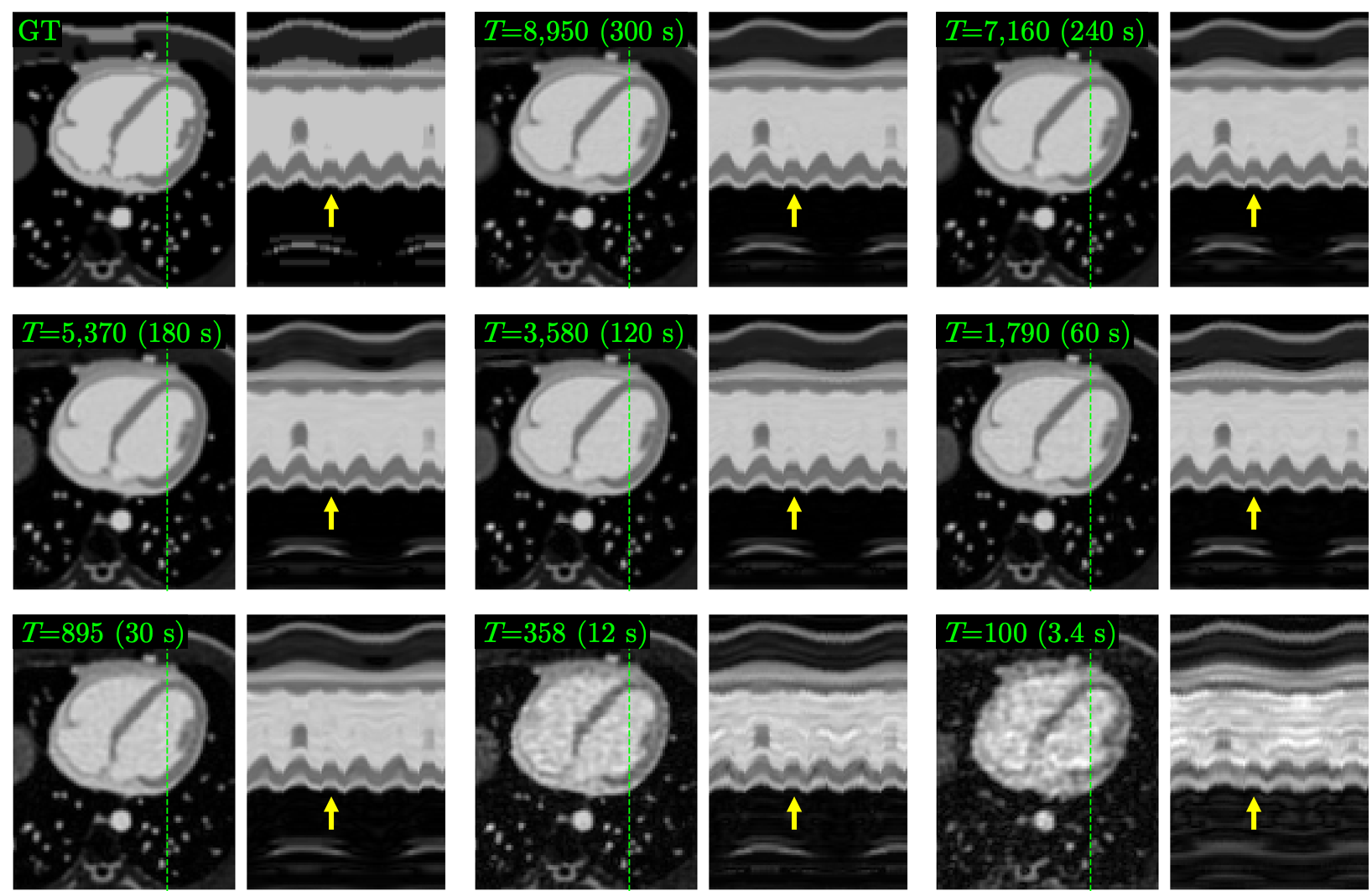}
    
    \caption{Representative images from MRXCAT reconstructed using ML-DIP. Time profiles along the dashed green lines are shown to the right of the images. Each x–t profile spans 100 frames. A simulated PVC beat is highlighted by the yellow arrow. Here, GT represents ground truth, $T$ represents the number of frames, and the numbers in the parentheses represent scan times, assuming a repetition time is similar to the ones reported for human studies. 
    }
    \label{fig:mrxcat}
\end{figure*}
\clearpage

\begin{figure*}
    \centering
    \includegraphics[width=0.6\linewidth]{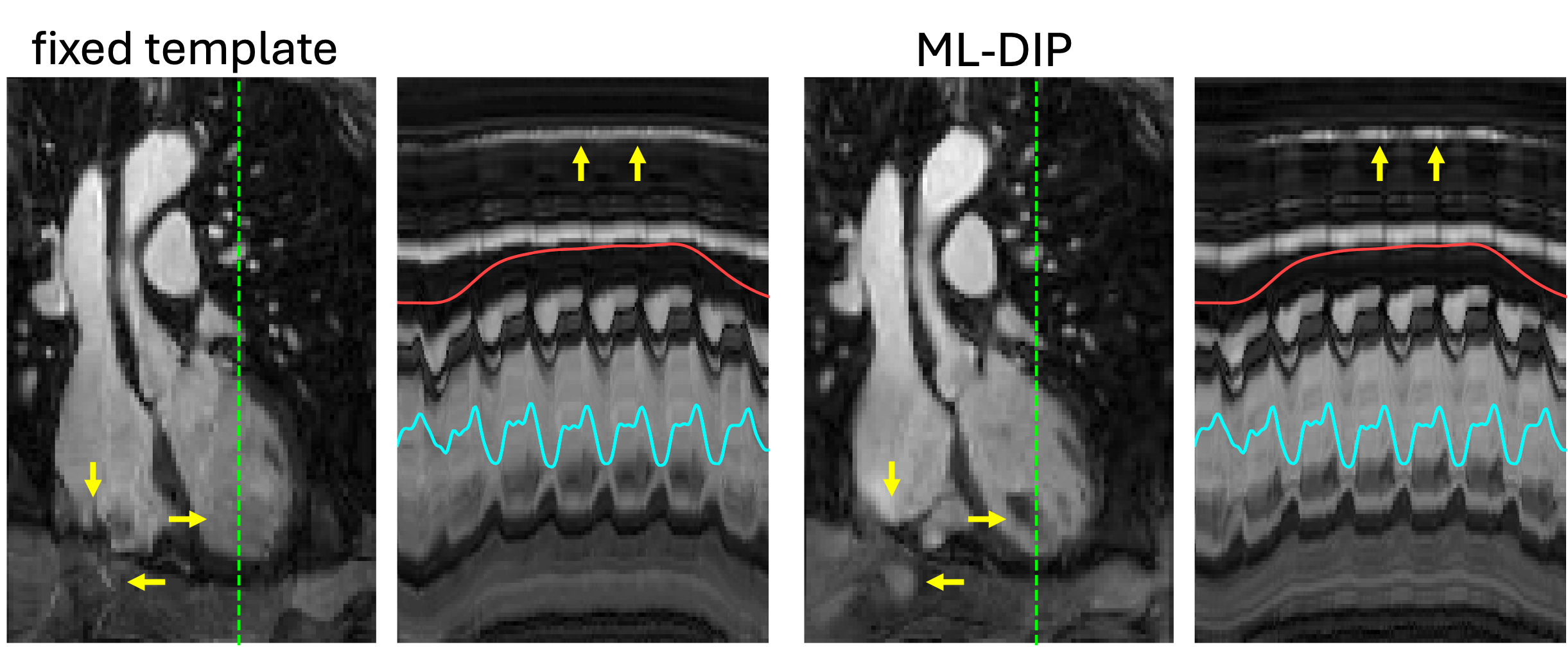}
    \caption{A representative coronal frame from the reconstructions using data from Vol.\ \#8. Space-time (x-t) profiles along the dashed green lines are shown to the right of the images. Each x–t profile spans 200 frames (6.6\,s). The reconstruction on the left uses a fixed (time-invariant) complex-valued template, while ML-DIP uses a frame-specific composite image. Some spatial and temporal details (yellow arrows) are lost or distorted in the image on the left.}
    \label{fig:fixed}
\end{figure*}
\clearpage

\clearpage


\clearpage


\clearpage


\begin{figure*}
    \centering
    \includegraphics[width=0.95\linewidth]{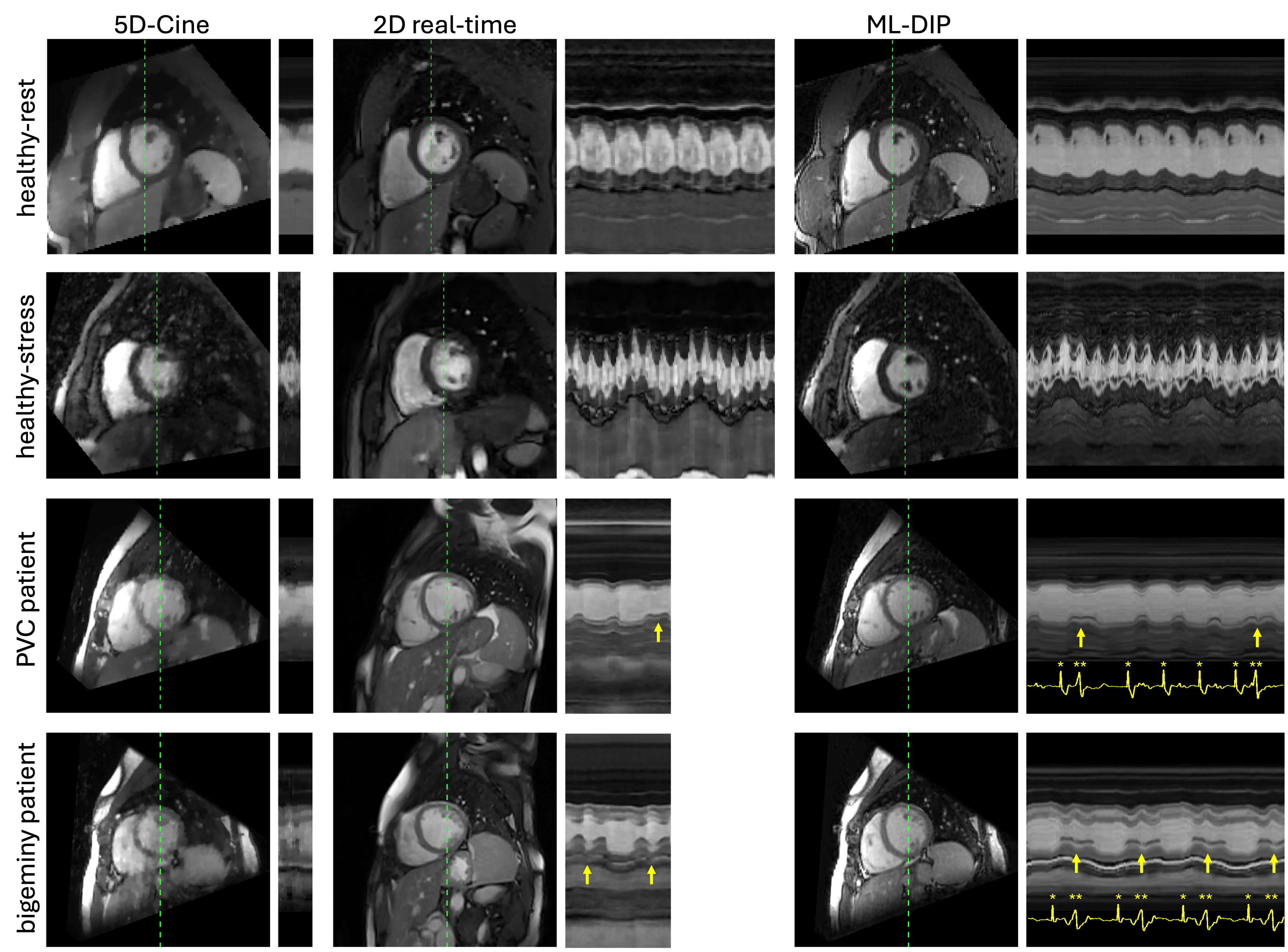}
    \caption{Representative short-axis images from 5D-Cine, 2D real-time, and ML-DIP for a healthy subject at rest, a healthy subject during exercise, and two patients. Space-time (x-t) profiles along the dashed green lines are shown to the right of the images. The ML-DIP x–t profiles span 200 frames (6.6\,s), while the x-t profiles span 3 to 6 seconds for 2D real-time cine and one cardiac cycle for 5D-Cine. The 3D reconstructions from 5D-Cine and ML-DIP were interpolated along the 2D plane defined by the 2D real-time acquisition. \textb{Yellow traces show the ECG signal that was synchronously collected with the 3D acquisition. R-waves from sinus and PVC beats are marked with `*' and `**,' respectively.  The yellow arrows indicate PVCs seen on the cine images.} Artifacts due to uncompensated motion are observed in 5D-Cine, especially in the second, third, and fourth rows. Because 2D and 3D scans were performed separately, minor shifts in subject position are seen in some cases. 
    }
    \label{fig:comparison}
\end{figure*}
\clearpage

\clearpage

\clearpage

\captionsetup{type=figure}
{\noindent\caption*{\textbf{Video S1.} A representative slice from MRXCAT reconstructed using ML-DIP. Here, $T$ represents the number of frames used during training. This video corresponds to \figref{mrxcat} of the manuscript and is available under the Ancillary files link on arXiv.}}

\vspace{0.5em}
\captionsetup{type=figure}
{\noindent\caption*{\textbf{Video S2.} Representative short-axis cine from 5D-Cine, 2D real-time cine, and ML-DIP for a healthy subject at rest, a healthy subject during exercise, a PVC patient with irregular rhythm, and a PVC patient with bigeminy. For the real-time cine (the two right columns), a pause of 1.5 s is inserted after each loop. This video corresponds to \figref{comparison} of the manuscript and is available under the Ancillary files link on arXiv.}}

\vspace{0.5em}
\captionsetup{type=figure}
{\noindent\caption*{\textbf{Video S3.} Coronal views from ML-DIP reconstruction from two different subjects. A healthy subject (left) and a PVC patient in bigeminy rhythm (right) are shown. Due to the strong relaxivity effects of ferumoxytol, the image on the left shows higher blood-myocardium contrast. Also, the cine on the right (collected with balanced steady-state free precession) shows stronger signal loss at the tissue-fat interface and is available under the Ancillary files link on arXiv.}}

\clearpage

\end{document}